\documentclass[twocolumn,showpacs,prb,
citeautoscript,amsmath,amssymb,floatfix,eqsecnum]{revtex4-2}
\usepackage{amsmath,amssymb,color}
\usepackage{bm}
\usepackage{graphicx}
\usepackage{psfrag}
\usepackage{bbold}
\usepackage{bbm}
\usepackage{hyperref}
\newcommand{\bd}{\bm}

\begin{document}

\title{Recursive 
algorithm for generating high-temperature expansions for spin systems \\ and the
chiral non-linear susceptibility}


\author{Andreas R\"{u}ckriegel}
  
\affiliation{Institut f\"{u}r Theoretische Physik, Universit\"{a}t
  Frankfurt,  Max-von-Laue Stra{\ss}e 1, 60438 Frankfurt, Germany}

\author{ Dmytro Tarasevych}
  
\affiliation{Institut f\"{u}r Theoretische Physik, Universit\"{a}t
  Frankfurt,  Max-von-Laue Stra{\ss}e 1, 60438 Frankfurt, Germany}

\author{Jan Krieg}
  

\affiliation{Institut f\"{u}r Theoretische Physik, Universit\"{a}t
  Frankfurt,  Max-von-Laue Stra{\ss}e 1, 60438 Frankfurt, Germany}

\author{Peter Kopietz}
  
\affiliation{Institut f\"{u}r Theoretische Physik, Universit\"{a}t
  Frankfurt,  Max-von-Laue Stra{\ss}e 1, 60438 Frankfurt, Germany}

\date{October 10, 2024}

 \begin{abstract}
We show that
the high-temperature expansion of the free energy and arbitrary 
imaginary-time-ordered
connected correlation functions
of quantum spin systems
can be recursively obtained from the exact 
renormalization
group flow equation for the generating functional of
connected spin correlation functions derived by Krieg and Kopietz [Phys. Rev. B {\bf{99}}, 060403(R) (2019)]. 
Our recursive algorithm can be explicitly written down in closed form
including all combinatorial factors.
We use our method to estimate critical temperatures of 
Heisenberg magnets from low-order 
truncations of the {\it{inverse}} spin susceptibility in the static limit.
We also 
calculate the connected correlation function
involving three different spin components
(chiral non-linear susceptibility)
of quantum Heisenberg magnets up to second order in the exchange couplings.
%

\end{abstract}


\maketitle

\section{Introduction}

High-temperature series expansions are a powerful numerical method to investigate 
lattice models for quantum magnetism or strongly correlated electrons \cite{Domb74,Oitmaa06}.
By numerically implementing graph theoretical and 
combinatorial algorithms it is nowadays possible  
to evaluate many terms in the high-temperature 
series \cite{Lohmann14,Mueller17,Hehn17,Gonzalez23,Pierre24}.
State-of-the art high-temperature expansions of 
quantum Heisenberg models include terms up to 
13th to 20th order for thermodynamic quantities and static correlation functions,
depending on the lattice type \cite{Gonzalez23,Pierre24}.
For classical spin models such as the Ising model the high-temperature series expansion 
can be pushed to even higher orders; for example, 
in
Ref.~[\onlinecite{Campostrini02}] the first $25$ terms 
of the high-temperature expansion of the 
zero-field susceptibility of
the nearest-neighbor Ising model on a cubic lattice 
have  been calculated. 
From the mathematical analysis of the
truncated series one can
obtain quantitative estimates for critical phenomena, such as phase boundaries, critical exponents, and critical temperatures.
High-temperature series expansions for
static two-spin correlation functions of quantum Heisenberg models 
have also been obtained~\cite{Hehn17}.

While sophisticated 
numerical algorithms for generating the high-temperature series expansions
for spin models are available \cite{Hehn17,Pierre24}, 
for the calculation of the first few terms in the high-temperature series expansion it 
would be  useful to have a simple algorithm which
can be readily implemented using symbolic manipulation software such as MATHEMATICA. 
Moreover, it would be useful to have an algorithm for calculating the high-temperature expansion of arbitrary dynamic correlation functions 
in a magnetic field. 
In this work we show that such a general algorithm can be
obtained from the formally exact functional renormalization group flow equations for the 
generating functional of connected time-ordered spin correlation functions of quantum spin systems derived
in Ref.~[\onlinecite{Krieg19}].
We then use our method to calculate the {\it{inverse}} momentum- and frequency-dependent spin susceptibility of a general class of quantum Heisenberg models in a magnetic field up to third order in the exchange couplings and use the result to estimate the critical temperatures
for specific models where benchmarks are available.
Finally, we use our algorithm to calculate the high-frequency behavior 
of the connected time-ordered correlation function 
of three different spin components
to second order in the exchange couplings.
This so-called {\it{chiral non-linear susceptibility}} \cite{Kappl23}
determines the quadratic response of the spin system to a time-dependent external magnetic field and exhibits a rather non-trivial frequency dependence which can in principle be measured experimentally. 
In an appendix,
we also compare our low-order series expansion to exact solutions of an analytically solvable toy model,
the Heisenberg trimer.

\section{Recursive algorithm  for  high-temperature 
expansions of
connected spin correlations}

In this section we derive a new recursive algorithm for generating the
high-temperature expansion of the free energy and arbitrary 
connected spin correlation functions of a general class of anisotropic quantum Heisenberg models with Hamiltonian
 \begin{equation}
 {\cal{H}} = \frac{1}{2} \sum_{ ij} \sum_{ ab }
J_{ ij}^{ ab} {S}^{a}_i {S}^{b}_j -  \sum_{i} \bd{H}_i \cdot \bd{S}_i,
 \label{eq:Hamiltonian}
 \end{equation} 
where
$J^{ab}_{ij}$ are arbitrary anisotropic exchange couplings,
$\bd{H}_i$  is  site-dependent  external magnetic field
in units of energy, and the spin-operators
${\bd{S}}_i$ are normalized such that 
$\bd{S}_i^2 = S ( S+1)$.
The subscripts $i$ and $j$  label the positions
$\bd{R}_i,$ and $\bd{R}_j$ of the spins
while  the superscripts 
$a$ and $b$ label their  Cartesian  components.
At this point the spin positions $\bd{R}_i$ are arbitrary, so that
our expressions given below are also valid for finite spin clusters
and systems where the spins are not located on a periodic lattice.
For simplicity, we only consider spin systems without single-site anisotropy so that
we can set
 \begin{equation}
 J^{ab}_{ ii} =0.
 \end{equation}
It is straightforward to generalize our method to include spin models with
finite single-site anisotropy.
We assume that the local magnetic fields are aligned with the $z$-axis so that 
the part of the Hamiltonian representing the Zeeman energy can be written as
 \begin{equation}
 {\cal{H}}_0 = - \sum_{i} \bd{H}_i \cdot \bd{S}_i = - \sum_i H_i S^z_i.
 \end{equation}
For arbitrary $\bd{H}_i$ we can locally rotate the coordinate system in spin space 
such that the rotated $z$ axis points into the 
direction of the local magnetic field. Note that such a rotation re-defines the exchange couplings, but since we consider arbitrary $J^{ab}_{ij}$ we do not 
lose any generality.

\begin{figure}[tb]
 \begin{center}
  \centering
\vspace{7mm}
 \includegraphics[width=0.4\textwidth]{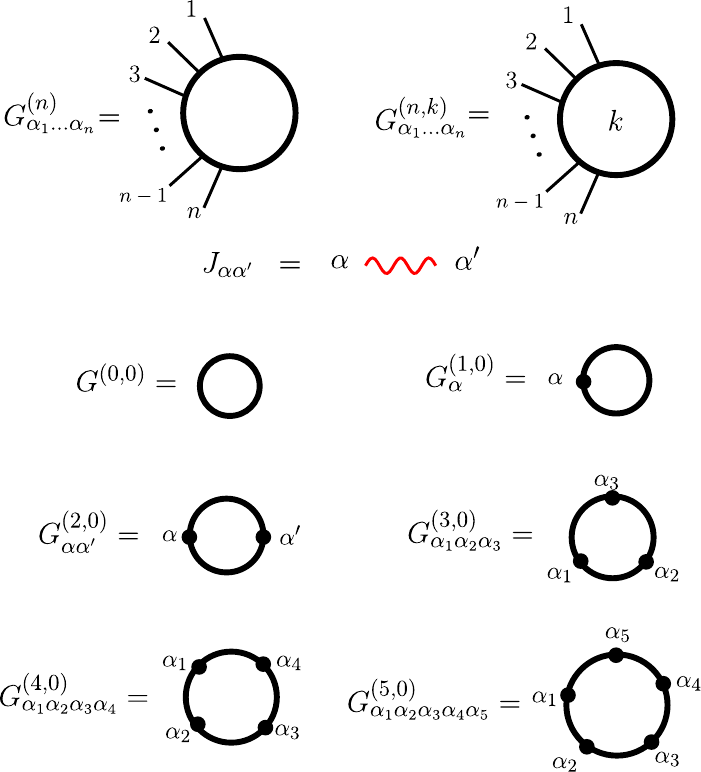}
   \end{center}
  \caption{%
Elements of our 
graphical representation of the perturbation series generated by the
iteration of Eq.~\eqref{eq:iterate}.  The exact connected $n$-spin correlation function 
$G^{(n)}_{\alpha_1 \ldots \alpha_n }$ 
is represented by a large  empty circle with $n$ labeled external legs.
A number $k$ inside the circle denotes the contribution 
$G^{(n,k)}_{\alpha_1 \ldots \alpha_n}$ of
 order $J^k$ to $G^{(n)}_{\alpha_1 \ldots \alpha_n }$. 
The exchange interaction is represented by a (red) wavy line.
Circles of varying size denote the corresponding connected spin correlation functions for $J=0$, where the dots represent the external labels.
These quantities correspond precisely to the generalized blocks 
introduced by Izyumov and Skryabin \cite{Izyumov88} who used the same graphical notation.
}
\label{fig:legend}
\end{figure}

\begin{figure}[tb]
 \begin{center}
  \centering
\vspace{7mm}
 \includegraphics[width=0.45\textwidth]{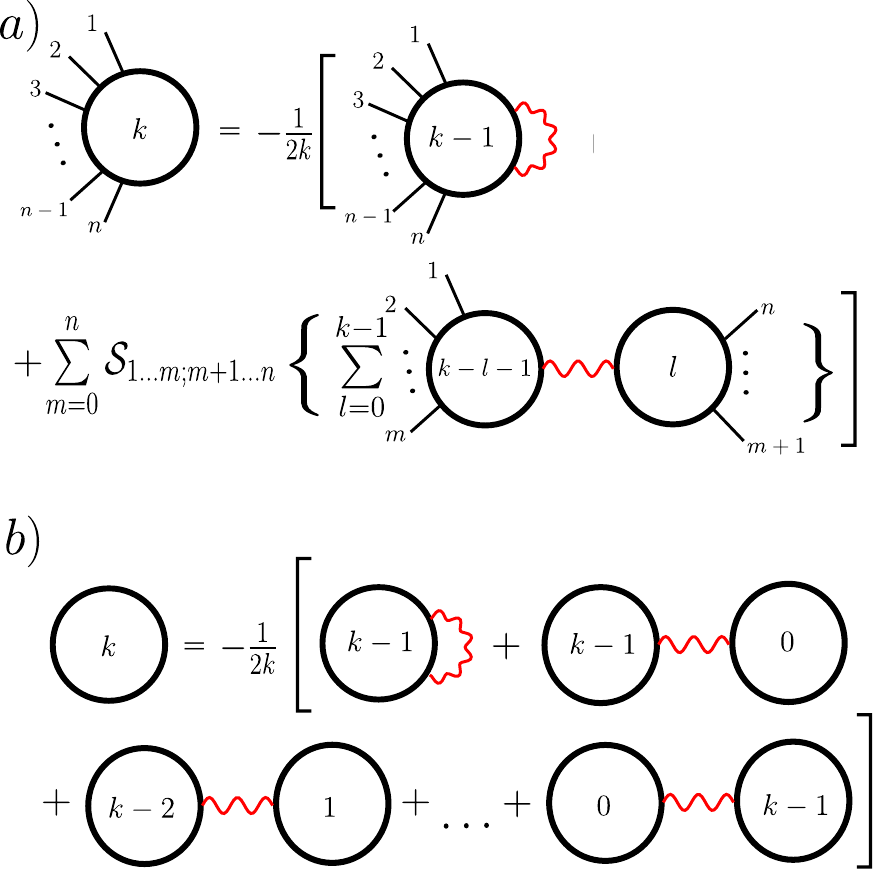}
   \end{center}
  \caption{%
(a) Graphical representation of the exact recursion formula \eqref{eq:iterate}
relating the connected $n$-spin correlation function of order $k$ 
in terms of a sum of connected 
spin correlation functions involving up to $n+2$ external legs
but at  most $k-1$ powers of the exchange couplings.
(b) Special case of (a) for the free energy which is represented by an empty circle without external legs. In this case we may omit the symmetrization operator because
there are no external labels to be symmetrized. 
Our recursion formula \eqref{eq:iterate} then relates the free energy
at order $k$ to the connected two-point function at order $k-1$ and a sum of products of one-point functions (magnetic moments) of order
$k-l-1$ and $l$  with $l =0, \ldots , k-1$.
}
\label{fig:recursion}
\end{figure}

Our algorithm is based on the formally exact functional renormalization group (FRG) flow equation for the generating functional of the connected spin correlation functions in imaginary time \cite{Krieg19}. For completeness, let us briefly outline the
derivation of this flow equation.
Following Ref.~[\onlinecite{Krieg19}], we replace the exchange couplings by a
deformation $J_{ij}^{ab} \rightarrow J^{ ab}_{ij, \Lambda}$ depending on a continuous
parameter $\Lambda$ such that $J_{ij, \Lambda =0}^{ ab} =0$ 
and $J_{ij, \Lambda=1}^{ ab} =J_{ij}^{ ab}$.
In this work we satisfy this boundary condition
via an interaction switch~\cite{Krieg19}
 \begin{equation}
 J_{ij, \Lambda}^{ ab} = \Lambda J_{ij}^{ ab}, \; \; \;  \Lambda \in [ 0, 1].
 \label{eq:switch}
 \end{equation}
In spite of its simplicity this deformation scheme has been useful in
several recent applications of the spin 
FRG approach \cite{Goll19,Goll20,Tarasevych21,Tarasevych22b,Rueckriegel24}.
Following Ref.~[\onlinecite{Krieg19}] we write
the deformed generating functional ${\cal{G}}_{\Lambda} [ \bd{h} ]$
of the connected time-ordered spin correlation functions in imaginary time $\tau$
as the trace of a time-ordered exponential,
  \begin{align}
 e^{ {\cal{G}}_{\Lambda} [ \bd{h} ]}
 = {} &  {\rm Tr} \Bigl\{  e^{ - \beta {\cal{H}}_{0 }} {\cal{T}} e^{ \int_0^{\beta} 
 d \tau    \sum_i \bd{h}_i ( \tau ) \cdot   {\bd{S}}_i ( \tau )  }
 \nonumber
 \\
 &  \times e^{ - \int_0^{\beta} d \tau
    \frac{1}{2} \sum_{ ij} \sum_{ab}
  {J}_{ ij, \Lambda}^{ab} {S}^{a}_i ( \tau )  {S}^{b}_j ( \tau ) }  \Bigr\}.
 \label{eq:Gcdef}
 \end{align}
Here $\beta$ is the inverse temperature, ${\cal{T}}$ denotes time ordering in imaginary time,
$\bd{h}_i ( \tau )$ are fluctuating source fields, 
and the time dependence of all operators is in the interaction picture with respect to  ${\cal{H}}_0$. By differentiating Eq.~\eqref{eq:Gcdef} with respect to the deformation parameter $\Lambda$ we 
obtain the  exact flow equation~\cite{Krieg19}
\begin{widetext}
\begin{equation}
\partial_{\Lambda}{\cal{G}}_{\Lambda} [ \bd{h} ]  
  = -
\frac{1}{2} \int_0^{\beta} d \tau 
 \sum_{ij,  ab} ( \partial_{\Lambda} {J}^{ ab}_{ij, \Lambda} ) 
 \Biggl[\frac{ \delta^2 {\cal{G}}_{\Lambda} [ \bd{h} ] }{\delta h_i^{a} ( \tau )   
\delta h_j^{b} ( \tau ) }
 + 
\frac{ \delta {\cal{G}}_{\Lambda} [ \bd{h} ] }{\delta h_i^{a} ( \tau ) }
\frac{ \delta {\cal{G}}_{\Lambda} [ \bd{h} ] }{\delta h_j^{b} ( \tau ) }
  \Biggr].
 \label{eq:flowW2}
 \end{equation}
This is equivalent to an infinite hierarchy of flow equations for 
the connected time-ordered $n$-spin correlation functions 
$G^{a_1 \ldots a_n}_{ i_1 
\ldots i_n,  \Lambda} ( \tau_1 , \ldots , \tau_n )$, which are defined via the
functional Taylor expansion of ${\cal{G}}_{\Lambda} [ \bd{h} ]  $:
 \begin{equation}
 {\cal{G}}_\Lambda [ \bd{h} ] = 
 {\cal{G}}_{\Lambda} [  0 ] + \sum_{ n=1}^{\infty} \frac{1}{ n!}
 \int_{0}^{\beta} d \tau_1 \ldots \int_0^{\beta} d \tau_n \sum_{ i_1 \ldots i_n} \sum_{ a_1 \ldots a_n} G^{a_1 \ldots a_n}_{ i_1 \ldots i_n, \Lambda } ( \tau_1 , \ldots , \tau_n )
 h^{a_1}_{ i_1 } ( \tau_1 ) \ldots h^{a_n}_{ i_n} ( \tau_n ).
 \end{equation}
 Omitting for simplicity the deformation label $\Lambda$
the hierarchy of flow equations implied by the functional differential equation~\eqref{eq:flowW2} can be written as \cite{Krieg19}
 \begin{align}
 & \partial_{\Lambda}    G_{ i_1 \ldots i_n }^{a_1 \ldots a_n } 
 ( \tau_1, \ldots,  \tau_n )   =   -  \frac{1}{2} \int_0^{\beta} d \tau 
 \sum_{ij, ab} ( \partial_{\Lambda} J^{ab}_{ij, \Lambda} )
 \biggl[G_{ i_1 \ldots i_n ij }^{a_1 \ldots a_n  ab  } 
 ( \tau_1, \ldots,  \tau_n, \tau , \tau  )
 \nonumber
 \\
 & + \sum_{ m=0}^n  {\cal{S}}_{ 1, \ldots, m; m+1, \ldots, n }
 \left\{
G_{ i_1 \ldots i_m i }^{a_1 \ldots a_m a  } 
 ( \tau_1, \ldots,  \tau_m, \tau   )
G_{ i_{m+1} \ldots i_n j }^{a_{m+1} \ldots a_n b  } 
 ( \tau_{m+1}, \ldots,  \tau_n, \tau   )
 \right\} \biggr].
 \label{eq:flowGn}
 \end{align}
 \end{widetext}
Here the symmetrization operator ${\cal{S}}_{ 1, \ldots, m; m+1, \ldots, n } \left\{ \ldots \right\}$ 
symmetrizes the expression in the curly braces with respect to the
exchange of all labels \cite{Kopietz10}. Explicitly, the action of this operator
on a function $f ( 1 , \ldots , n ) $ which is already symmetric with respect to the first $m$ labels $1 , \ldots, m$ and the last $n-m$ labels $m+1, \ldots , n$
is
 \begin{align}
  &   {\cal{S}}_{ 1, \ldots, m; m+1, \ldots, n }
 \left\{ f ( 1 , \ldots , n ) \right\}
 \nonumber
 \\ 
 = {} &
 \frac{1}{ m! ( n-m)! } \sum_{P}  f ( P_1 , \ldots , P_n ),
 \label{eq:symdef}
 \end{align}
where $\sum_{P}$ is the sum of all $n!$ permutations 
$(P_1 , \ldots , P_n )$
of  $(1, \ldots , n )$.
Note that the number of terms generated by ${\cal{S}}_{ 1, \ldots, m; m+1, \ldots, n } \left\{ \ldots \right\}$ is given by the binomial coefficient
 \begin{equation}
  \begin{pmatrix} n \\ m \end{pmatrix}  =
  \frac{ n!}{  m! (n-m)!}.
  \end{equation}
A graphical representation of Eq.~\eqref{eq:flowGn} can be found in 
Fig.~1 of Ref.~[\onlinecite{Krieg19}].
At this point 
it is useful to introduce multi-labels $\alpha = ( i , \tau , a )$
representing
all parameters which are necessary to specify 
the degrees of freedom, i.e., lattice site, imaginary time, and field component.
Our FRG flow equation \eqref{eq:flowGn} can then be written as
 \begin{align}
   & \partial_{\Lambda}    G^{(n)}_{ \alpha_1 \ldots \alpha_n }
    =   -  \frac{1}{2}  \int_{\alpha} \int_{ \alpha^{\prime}}
  \left[ ( \partial_{\Lambda} \mathbf{J}_{\Lambda} ) \right]_{\alpha \alpha^{\prime}}
  \biggl[
  {G}^{(n+2)}_{ \alpha \alpha^{\prime}\alpha_1 \ldots \alpha_n } 
 \nonumber
 \\
 &  + \sum_{ m=0}^n  {\cal{S}}_{ 1, \ldots, m; m+1, \ldots, n }
 \left\{
 {G}^{(m+1)}_{\alpha \alpha_1 \ldots \alpha_m } 
 {G}^{(n-m+1)}_{ \alpha^{\prime} \alpha_{m+1} \ldots \alpha_n } 
 \right\} \biggr],
 \label{eq:flowGn2}
 \end{align}
where $\int_{\alpha} = \int_0^{\beta} d \tau \sum_{ i a }$ and
$\mathbf{J}_{\Lambda}$ is a matrix in the multi-labels with matrix elements
\begin{equation}
 [ \mathbf{J}_{\Lambda} ]_{ \alpha = ( i  \tau a), \alpha^{\prime} = ( j \tau^{\prime} b )}
  = \delta ( \tau - \tau^{\prime} ) J^{ab}_{ ij , \Lambda}.
 \end{equation}
%
The hierarchy \eqref{eq:flowGn2} of flow equations
can now be used to generate a systematic expansion
of the connected spin correlation functions in powers of the exchange couplings, which amounts to a high-temperature expansion.
The crucial observation is that with the interaction switch deformation scheme 
\eqref{eq:switch}
an expansion in powers of $\Lambda$ amounts to an expansion in powers of $J/T$ \cite{Rueckriegel24}. 
To generate the high-temperature expansion, we simply substitute 
the Taylor expansion 
 \begin{equation}
 G^{(n)}_{ \alpha_1 \ldots \alpha_n , \Lambda}
  = \sum_{ k = 0}^{\infty} \Lambda^{k} G^{(n,k)}_{ \alpha_1 \ldots \alpha_n}
 \end{equation}
into  the flow equation~\eqref{eq:flowGn2} and compare  powers of $\Lambda$
on both sides.  This yields the recursion relation,
 \begin{align}
 &  G^{(n,k)}_{  \alpha_1 \ldots \alpha_n } = 
 -  \frac{1}{2 k }  \int_{\alpha} \int_{ \alpha^{\prime}}
  {J}_{\alpha \alpha^{\prime}}
  \Biggl[
   {G}^{(n+2,k-1)}_{ \alpha \alpha^{\prime} \alpha_1 \ldots \alpha_n }  +
  \nonumber
  \\
  &  \hspace{-5mm}    \sum_{ m=0}^n    {\cal{S}}_{ 1, \ldots, m; m+1, \ldots, n }
 \biggl\{ \sum_{l=0}^{k-1}   G^{(m+1,l)}_{ \alpha \alpha_1 \ldots \alpha_m }
  {G}^{(n-m+1, k-l-1)}_{ \alpha^{\prime} \alpha_{m+1} \ldots \alpha_n }
 \biggr\}  \Biggr],
 \label{eq:iterate}
  \end{align}
%
which is the central result of this work.
Keeping in mind that for $\Lambda =1$ we recover our original Heisenberg model,
the high-temperature expansion of the physical connected $n$-spin correlation
function is 
\begin{equation}
 G^{(n)}_{ \alpha_1 \ldots \alpha_n }
  = \sum_{ k = 0}^{\infty} G^{(n,k)}_{ \alpha_1 \ldots \alpha_n}.
 \end{equation} 
We emphasize that the recursion relation \eqref{eq:iterate} provides one single, compact, and fully analytical formula valid for any term of the perturbation series of any imaginary-time-ordered $n$-spin correlation function.
To define a graphical representation of
the recursion relation  we introduce the graphical elements
shown in Fig.~\ref{fig:legend}.
With this notation our recursion 
recursion relation \eqref{eq:iterate}
can be represented graphically as shown
in Fig.~\ref{fig:recursion}.

The  exact recursion relation 
\eqref{eq:iterate} expresses the connected $n$-point function $G^{(n,k)}$ at order $(J/T)^k$
in terms of the connected $m$-point functions  $ G^{(m,l)}$ with  $1 \leq  m  \leq n+2$ and $0 \leq  l \leq k -1$. 
Each iteration lowers the order $k$ of the expansion
by one but raises the number $n$ of external legs 
by two. After $k$ iterations all terms generated by this recursion
can be expressed in terms of the connected correlation functions 
$G^{(m,0)}$ of an isolated spin, i.e.,  the {\it{generalized blocks}}
introduced by Izyumov and Skryabin \cite{Izyumov88}.
Following their notation, we represent the generalized blocks graphically
by circles of varying size with black dots representing the external labels as shown
in Fig.~\ref{fig:legend}.
Given the fact that for coinciding lattice sites $J^{ab}_{ii} =0$,
we see that the contribution from the term  
involving $n + 2k$ external legs  vanishes,
\begin{equation}
  \int_{\alpha} \int_{ \alpha^{\prime}} {J}_{\alpha \alpha^{\prime}}
  {G}^{(n+2k,0)}_{ \alpha \alpha^{\prime} \alpha_1 \ldots \alpha_n } =0.
 \end{equation}
By iteration we  can thus  express an arbitrary connected $n$-point function $G^{(n,k)}$ at order $ (J/T)^k$  in terms of 
the connected correlation functions
$G^{(m,0)}$ involving  up to $m = n + 2k -1$ 
components of an  isolated spin. 
Keeping in mind that for fixed $m$ the symmetrization operator ${\cal{S}}_{ 1 , \ldots , m; m+1 , \ldots , n }$ in the second line of Eq.~\eqref{eq:iterate} generates 
$ n! /( m! ( n-m)!) $ different terms corresponding to distinct permutations of the external labels, the total number of terms generated by the nested sum
in  the second line of Eq.~\eqref{eq:iterate} is
 \begin{equation}
  \sum_{ m=0}^{n}  
  \sum_{ l =0}^{k-1}
 \begin{pmatrix} n \\ m \end{pmatrix}  =    2^n k.
 \end{equation}
All combinatorial factors are automatically generated by the symmetrization operator   ${\cal{S}}_{ 1 , \ldots , m; m+1 , \ldots , n }$ so that the 
 lowest few terms in the high-temperature expansion of the free energy and 
arbitrary connected correlation functions can be easily obtained purely 
algebraically
without additional combinatorial considerations.
Of course, the explicit calculation of 
high orders in this expansion requires a numerical implementation of the recursion relation \eqref{eq:iterate} which is beyond the scope of this work. In fact, at this point it is not clear
whether
our recursive algorithm offers any computational advantages over the
established methods \cite{Domb74,Oitmaa06} for generating the high-temperature expansions of Ising or Heisenberg models.

\section{Truncated high-temperature expansions}
\label{sec:trunc}

The first few terms in the  high temperature expansions of the free energy and  
the connected $n$-point correlation functions 
for a general spin Hamiltonian of the form~\eqref{eq:Hamiltonian}
can be obtained analytically by 
straightforward iteration of our recursion relation \eqref{eq:iterate}.
To that end, 
it is convenient to transform the imaginary-time correlation functions to frequency space. 
We normalize the Fourier transform of the connected $n$-spin correlation function as follows \cite{Goll19},
 \begin{align}
  & G_{ i_1 \ldots i_n }^{ a_1 \ldots a_n} ( \tau_1 , \ldots , \tau_n ) 
 = \frac{1}{\beta^n} \sum_{ \omega_1 \ldots \omega_n }
 e^{ - i ( \omega_1 \tau_1 + \ldots + \omega_n \tau_n ) }
 \nonumber
 \\
&  \times  \beta \delta_{ \omega_1 + \ldots + \omega_n, 0 } 
 {G}_{ i_1 \ldots i_n }^{ a_1 \ldots a_n} ( \omega_1 , \ldots , \omega_n ), \hspace{9mm}
 \label{eq:frequencycon}
 \end{align}
where we have used time-translational invariance to factorize  
a frequency-conserving Kronecker-$\delta$.
For two-spin correlation functions we suppress the redundant second frequency label.
In the atomic limit $J \rightarrow 0$ the corresponding 
correlation functions are diagonal in the site index,
 \begin{align}
 & \lim_{ J \rightarrow 0}  {G}_{ i_1 \ldots i_n }^{ a_1 \ldots a_n} ( \omega_1 , \ldots , \omega_n )
\nonumber
 \\
 = {} &  \delta_{i_1  i_2} \delta_{ i_2 i_3 } \ldots \delta_{ i_{n-1} i_n }
 g_{i_1}^{ a_1 \ldots a_n} ( \omega_1 , \ldots , \omega_n ),
 \end{align}
where the site dependence of $g_{i_1}^{ a_1 \ldots a_n} ( \omega_1 , \ldots , \omega_n )$ is due to the inhomogeneous
external magnetic field $H_i$. For a homogeneous field $H_i = H$ the
connected single-spin correlation functions
$g_{i_1}^{ a_1 \ldots a_n} ( \omega_1 , \ldots , \omega_n ) =   
g^{ a_1 \ldots a_n} ( \omega_1 , \ldots , \omega_n )$  are independent
of the site labels. 
Following Ref.~[\onlinecite{Izyumov88}] we refer to the functions
$g^{ a_1 \ldots a_n} ( \omega_1 , \ldots , \omega_n )$ as {\it{generalized blocks}}
(or simply {\it{blocks}})
and represent them graphically as shown in the lower part of Fig.~\ref{fig:legend}.
Explicit expressions for the blocks involving up to $n=4$ spin components 
have already been derived a long time ago by Vaks, Larkin, and Pikin \cite{Vaks68,Vaks68b},
and can also be found in the textbook~[\onlinecite{Izyumov88}].
A recursive algorithm for calculating the blocks for arbitrary $n$ has been constructed 
in Ref.~[\onlinecite{Goll19}] (see also Ref.~[\onlinecite{Halbinger23}]).
In Appendix~\ref{app:single_spin} we explicitly give the  blocks involving up to  $n=5$ spin components. As will become evident below, the five-spin blocks are necessary to obtain the
two-point function to order $J^3$ and the three-point function to order $J^2$.

\subsection{Free energy and magnetization}

\begin{figure}[htb]
 \begin{center}
  \centering
\vspace{7mm}
 \includegraphics[width=0.45\textwidth]{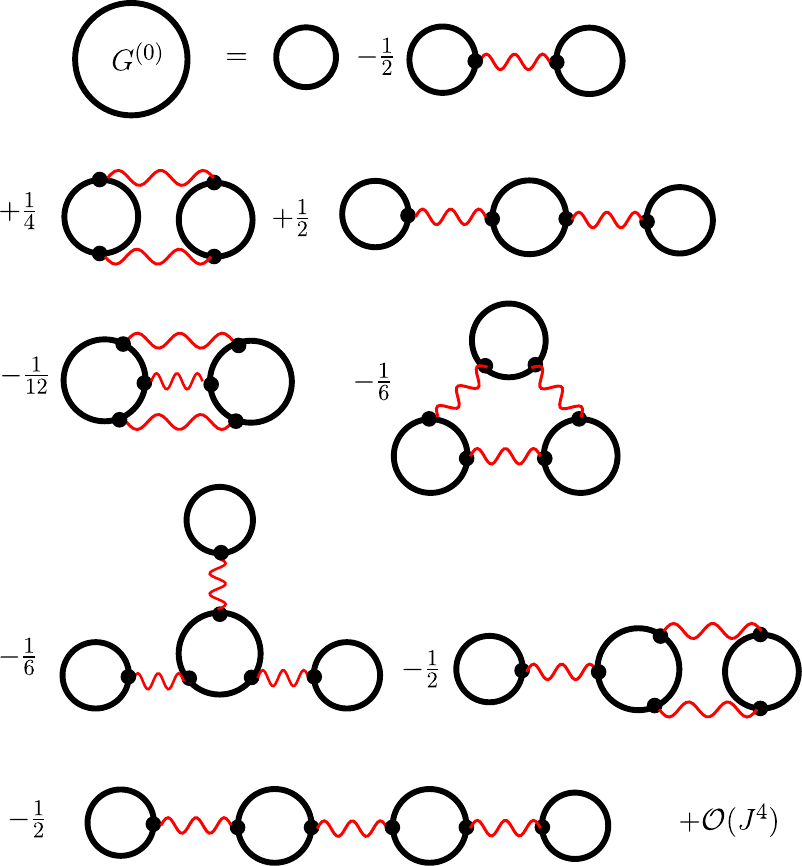}
   \end{center}
  \caption{%
Diagrammatic representation of the high-temperature series for the
negative free energy $G^{(0)} = - \beta F$ in units of temperature
up to order $J^3$. The symbols are defined in Fig.~\ref{fig:legend}.
}
\label{fig:freeenergy}
\end{figure}

Let us now calculate the high-temperature expansion of the negative free energy 
$ G^{(0)} = - \beta F $ in units of temperature
up to third order in $J$. To generate this expansion we set
$n=0$ in Eq.~\eqref{eq:iterate} and iterate up to order $k=3$, which can be easily done analytically. 
A graphical representation of the terms in this expansion
up to order $J^3$
is shown in Fig.~\ref{fig:freeenergy}. 
These graphs represent the following mathematical expression,
 \begin{align}
 G^{(0)} = {} &  \sum_i B ( \beta H_i )    - \frac{\beta}{2}  \sum_{ij}  J_{ij}^{zz} m_i m_j
 \nonumber
 \\
 & +  G^{(0,2)} + G^{(0,3)} + {\cal{O}} ( J^4 ),
 \label{eq:expansionfree}
 \end{align}
where
 \begin{equation}
 B ( y ) = \ln \left[ \frac{\sinh [ ( S+ 1/2) y ]}{\sinh ( y/2) } \right]
 \end{equation}
is the primitive integral of the spin-$S$ Brillouin function
 \begin{equation}
 b ( y)  = \left( S + \frac{1}{2} \right) \coth \left[ \left( S + \frac{1}{2} 
   \right) y \right] - \frac{1}{2} \coth \left[ \frac{y}{2} \right],
   \label{eq:bdef}
 \end{equation} 
which gives the local magnetic moments for vanishing exchange interaction,
 \begin{equation}
 m_i = b ( \beta H_i ).
 \end{equation}
For later reference we note that for small $y = \beta H$ the Brillouin function 
has the expansion
 \begin{equation}
 b( y ) = b_1 y + \frac{b_3}{3!} y^3 + {\cal{O}} ( y^5 ),
 \end{equation}
with
 \begin{subequations}
 \label{eq:b13def}
 \begin{align}
 b_1 & =     \frac{ (2S+1)^2 -1}{12} = \frac{ S ( S+1)}{3},
 \label{eq:b1def}
 \\
 b_3 & = - \frac{ (2S+1)^4 -1}{120}  = - \frac{6}{5} b_1 \left(
 b_1 + \frac{1}{6} \right).
 \label{eq:b3def}
 \end{align}
 \end{subequations}
The Cartesian components of the connected two-spin correlation function
of an isolated spin  
are in frequency space
 \begin{subequations}
 \label{eq:gabCartesian}
  \begin{align}
  g^{xx}_{ i} ( \omega ) & = g^{yy}_i ( \omega ) = \frac{ m_i H_i}{ H_i^2 + \omega^2 },
 \\
 g^{xy}_i ( \omega ) &  = - g_i^{yx} ( \omega ) = - \frac{ m_i \omega }{ H_i^2 + \omega^2 },
 \\
   g^{zz}_i ( \omega ) & =  \beta \delta_{\omega,0} b^{\prime} ( \beta H_i ),
 \label{eq:gzz}
 \end{align}
 \end{subequations} 
where $b^{\prime} ( y ) $ is the derivative of the Brillouin function.
For finite magnetic field it is more convenient to work with the spherical spin components,
 \begin{subequations}
 \label{eq:spherical}
 \begin{align}
 S_i^{+} & =  \frac{ S_i^x + i S_i^{y} }{\sqrt{2} },
 \\
 S_i^{-} & =  \frac{ S_i^x - i S_i^{y} }{\sqrt{2} },
 \\
 S_i^{0} & = S_i^z.
 \end{align}
 \end{subequations}
In this basis the Heisenberg equations of motion for the spin components decouple, 
 \begin{equation}
 \partial_{\tau} S_i^p ( \tau ) = - p H_i S_i^p ( \tau ), \; \;  \mbox{where $p =+,-,0$},
 \label{eq:eom}
 \end{equation} 
implying
 \begin{equation}
 S_i^p ( \tau ) = e^{ -p H_i \tau} S_i^p.
 \end{equation}
In frequency space the transverse connected two-spin correlation function in the spherical basis is therefore
 \begin{equation} \label{eq:gpp} 
   g^{+-}_i ( \omega ) = g^{-+}_i ( - \omega ) = \frac{m_i }{ H_i - i \omega }.
   \end{equation}
Note that for $H_i \rightarrow 0$  the functions $g_i^{zz} ( \omega )$, $g_i^{+-} ( \omega )$, and
$g_i^{-+} ( \omega )$  approach the same limit 
 \begin{equation}
 \lim_{ H_i \rightarrow 0} g^{ p \bar{p}}_i ( \omega ) =  g_0 ( \omega ) = 
\beta \delta_{\omega , 0 } b_1,
 \label{eq:g0def}
  \end{equation}
where $\bar{p} = - p$.
With the above  notation the second-order correction to the negative free energy in units of temperature  can be written as
 \begin{align}
 G^{(0,2)} = {} & 
\frac{1}{4} \sum_{ij} \sum_{ ab cd}   J^{a b}_{ij} J^{cd}_{ij} 
 \sum_{\omega} g^{a c}_i ( \omega ) g^{db}_j ( \omega )
 \nonumber
 \\
 & + 
 \frac{\beta}{2} \sum_{ijk} \sum_{ab} J^{za}_{ij}     J^{bz}_{jk} 
 g^{ab}_j (0) m_i m_k.
 \end{align}
The  third-order correction $G^{(0,3)}$ is rather lengthy and is explicitly 
given in Eq.~\eqref{eq:G03} of  Appendix~\ref{app:hightemp}.
There, we also explicitly evaluate the high-temperature series 
of the free energy of an arbitrary spin-$S$ Heisenberg magnet with $ H_i = 0 $ to order $ J^3 $ [see Eq.~\eqref{eq:F_third_order}].

In general, the contribution $G^{(0,k)}$ to the dimensionless free energy
of order $J^k$ depends on
generalized blocks $g^{a_1 \ldots a_l}_{ i} ( \omega_1 , \ldots , \omega_l )$ 
with up to $l=k$ external legs.
From the high-temperature expansion of the  free energy $F = - T G^{(0)}$
we can obtain the corresponding expansion of
the local magnetization $M_i$ 
to the same order 
in $J$ by taking a derivative with respect to the magnetic field,
 \begin{align}
 M_i  \equiv {} &  -  \frac{ \partial F }{\partial H_i } =  m_i - 
 \frac{\partial m_i}{\partial H_i }
\sum_j J^{zz}_{ij}
 m_j    
 \nonumber
 \\
 &+  
 \frac{ 1}{\beta} \frac{ \partial G^{(0,2)}}{\partial H_i }
+  \frac{ 1}{\beta} \frac{ \partial G^{(0,3)}}{\partial H_i }
+ {\cal{O}} ( J^4) \; . 
 \label{eq:MH}
 \end{align}
This expression gives the magnetic equation of state 
$M_i = M_i (  H  , T )$ of
spin models with Hamiltonian of the type \eqref{eq:Hamiltonian} 
at high temperatures up to order $J^3$. 
The explicit evaluation of Eq.~\eqref{eq:MH} for specific models such as the Kitaev-Heisenberg-$\Gamma$ model \cite{Rau14}, effective spin models for altermagnets \cite{Smejkal23,Gaitan24}
of other spin models of current interest
is straightforward but tedious and is not the subject of this work.

\subsection{Dynamic spin susceptibility}

\label{sec:dynamic_spin_susceptibility}

\begin{figure}[b]
 \begin{center}
  \centering
\vspace{7mm}
 \includegraphics[width=0.45\textwidth]{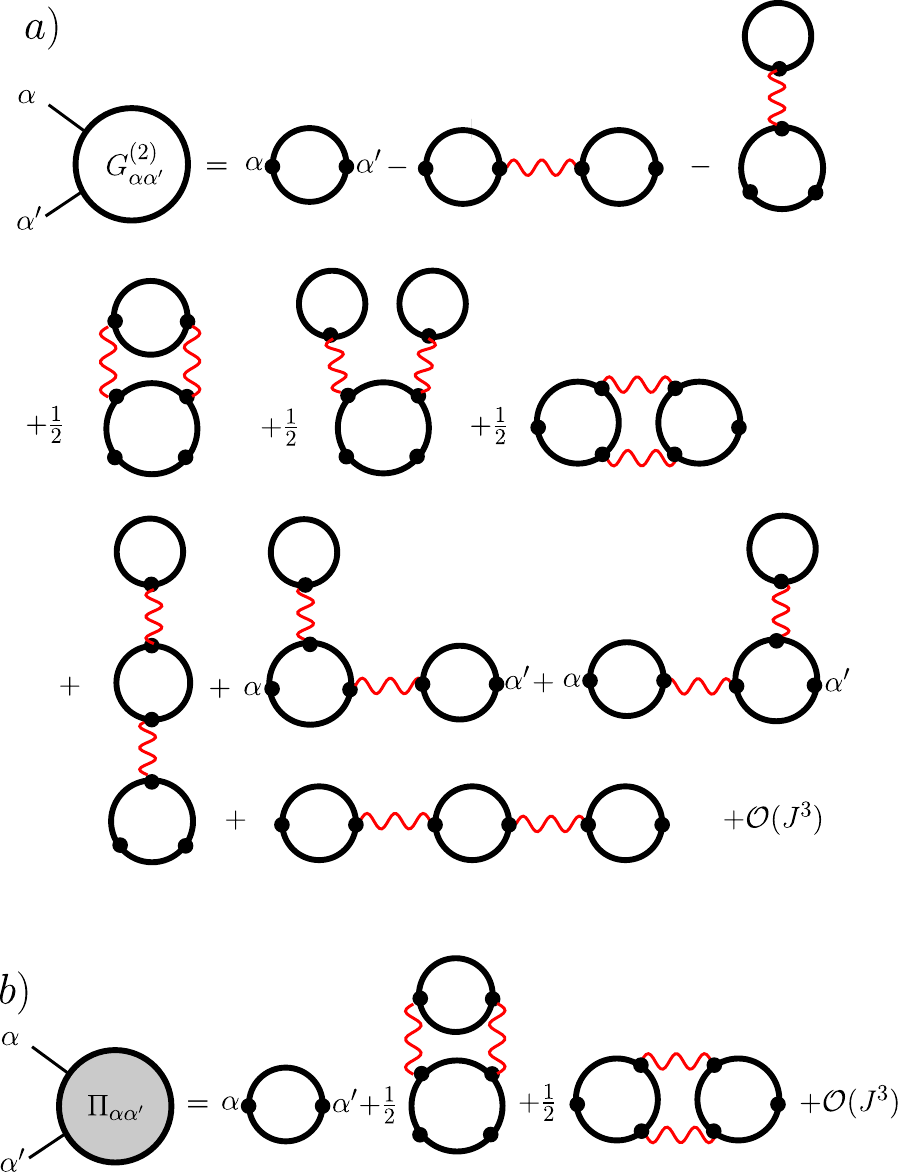}
   \end{center}
  \caption{%
(a) Expansion of the connected two-spin correlation function 
$G^{(2)}_{ \alpha \alpha^{\prime}}$
up to second order in the
exchange interaction.
(b) Expansion of the interaction-irreducible part 
$\Pi_{\alpha \alpha^{\prime}}$
of the 
connected two-spin correlation function
up to second order in the exchange interaction for vanishing magnetic field.
}
\label{fig:G2expansion}
\end{figure}

\begin{figure}[htb]
 \begin{center}
  \centering
\vspace{7mm}
 \includegraphics[width=0.45\textwidth]{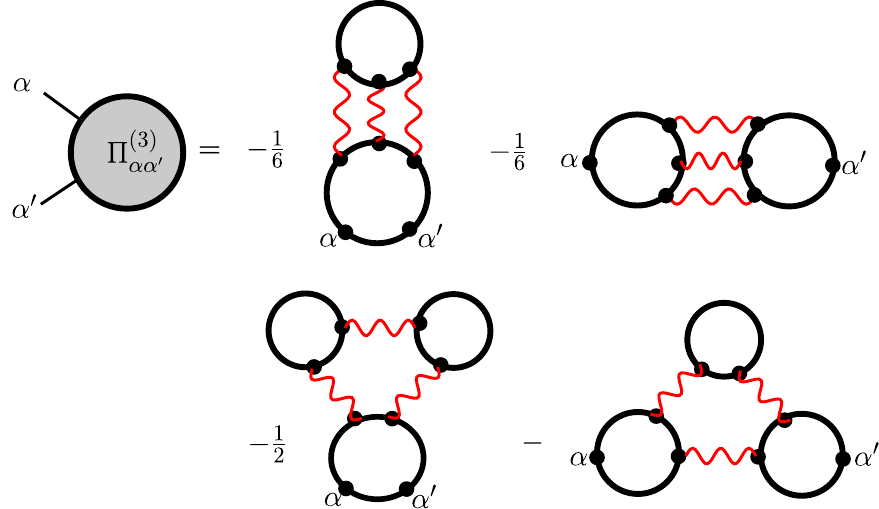}
   \end{center}
  \caption{%
Diagrammatic representation of the third-order contribution
 $\Pi^{(3)}_{\alpha \alpha^{\prime}}$
to the interaction-irreducible susceptibility.
}
\label{fig:Pi34}
\end{figure}

Next, we calculate  the high-temperature expansion of the connected two-spin correlation function up to order~$J^3$,
\begin{align}
 G^{  ab }_{ij} ( \omega )   = {} & \delta_{ij}  g_i^{ab} ( \omega ) + 
 G^{ ab (1) }_{ij} ( \omega ) +  G^{   ab (2)}_{ij} ( \omega ) 
 \nonumber
 \\
 & +  
 G^{ ab (3) }_{ij} ( \omega )  + {\cal{O}}   ( J^4 ).
 \label{eq:G2exp}
 \end{align}
The terms contributing to the expansion \eqref{eq:G2exp} 
up to order $J^2$ are shown diagrammatically in Fig.~\ref{fig:G2expansion} (a).
The mathematical expression for the first-order correction is
 \begin{align}
 G^{ ab (1) }_{ij} ( \omega ) = {} &
- \sum_{ cd}
 g_i^{a c} ( \omega )    J^{cd}_{ ij}    g_j^{d b} ( \omega )
 \nonumber
 \\
 & -  \delta_{ij} \sum_{ k} \sum_c     g_i^{ ab c} ( \omega , - \omega , 0 ) J_{ ik}^{cz}   m_k. \hspace{7mm}
 \label{eq:Gab1}
 \end{align}
The expressions for the 
seven second-order diagrams shown in Fig.~\ref{fig:G2expansion} (a)
are given in Eq.~\eqref{eq:G22} of Appendix~\ref{app:hightemp}.
The explicit evaluation of these expressions is rather tedious. To reduce the complexity, let us focus here
on the case of vanishing magnetic field where all tadpole diagrams vanish
and the spherical components of the two-point function
are given by $g_0 ( \omega ) = \beta \delta_{\omega , 0 } b_1$ [see Eq.~\eqref{eq:g0def}].
Moreover, in this limit only two of the diagrams in Fig.~\ref{fig:G2expansion} (a)
can be separated into two parts by cutting a single interaction line; these are the
interaction-reducible diagrams. Formally, the interaction-irreducible part
$\Pi_{\alpha \alpha^{\prime}}$ of the two-spin correlation function
$G_{\alpha \alpha^{\prime}} \equiv G^{(2)}_{\alpha \alpha^{\prime}}$ can be defined by setting
 \begin{equation}
 {G}_{\alpha \alpha^{\prime}}  = 
  \left[ \mathbf{\Pi} ( 1 +  \mathbf{J}\mathbf{\Pi} )^{-1} \right]_{\alpha \alpha^{\prime}}
 = \left[  \mathbf{\Pi}^{-1} + \mathbf{J} \right]^{-1}_{\alpha \alpha^{\prime}},
 \label{eq:GPi}
 \end{equation}
where all quantities on the right-hand side are matrices in the multi-labels. 
In Fig.~\ref{fig:G2expansion} (b) we show 
the diagrams contributing to the interaction-irreducible part
$\Pi_{\alpha \alpha^{\prime}}$ of the susceptibility up to second order in $J$, while
the third-order diagrams  
are shown in Fig.~\ref{fig:Pi34}.
For simplicity, let us focus now on an isotropic Heisenberg magnet where $J^{ab}_{ij}
 = \delta^{ab} J_{ij}$ on a Bravais lattice with $N$ sites. Then the matrix equation \eqref{eq:GPi}
can be diagonalized in momentum space so that the Fourier transform of the two-spin 
correlation function can be written as
 \begin{equation}
 G( \bd{k} ,  \omega  ) = \frac{\Pi ( \bd{k} ,  \omega  )}{ 1 + J_{\bd{k} } \Pi ( \bd{k} ,  \omega ) } =
 \frac{1}{ J_{ \bd{k} } + \Pi^{-1} ( \bd{k} ,  \omega ) },
 \end{equation}
where
 \begin{equation}
  {J}_{ {\bd{k}} }   =  \frac{1}{N} \sum_{ij} 
  e^{ - i \bd{k} \cdot ( \bd{R}_i - \bd{R}_j ) } {J}_{ij}
 \end{equation}
is the Fourier transform of the exchange couplings.
Evaluating the diagrams  in Figs.~\ref{fig:G2expansion} and \ref{fig:Pi34}
for vanishing magnetic field we obtain 
 \begin{equation}
  \Pi ( \bd{k} , \omega ) = \beta \delta_{\omega,0} b_1 +
   \Pi^{(2)} ( \bd{k} , \omega ) + \Pi^{(3)} ( \bd{k} , \omega )
    + {\cal{O}} ( J^4 ),
  \end{equation}
where the second-order term is
\begin{widetext}
 \begin{align}
 \Pi^{(2)}  ( \bd{k} , \omega  ) = {} & 
   \frac{1}{2 \beta N} \sum_{\bd{q}} J^2_{ \bd{q} }  
  \sum_{ \omega^{\prime}} g_0 (  \omega^{\prime} )
  \left[ 
   2 g_0^{+-zz} ( \omega^{\prime} , - \omega^{\prime} , \omega , - \omega ) 
  + g_0^{zzzz} ( \omega^{\prime} , - \omega^{\prime} , \omega , - \omega ) 
 \right]
 \nonumber
 \\
 & +   \frac{1}{\beta N} \sum_{\bd{q}} J_{ \bd{q} } J_{ \bd{q} + \bd{k} }  
  \sum_{ \omega^{\prime}} 
 g_0^{+-z} ( \omega^{\prime} , - \omega - \omega^{\prime} , \omega )
 g_0^{+-z} ( \omega + \omega^{\prime} ,  - \omega^{\prime} , - \omega ).
 \label{eq:Pi2int}
 \end{align}
Here the zero-field limits of the generalized blocks are denoted by $g_0 ( \omega )$,
$g_0^{+-z} ( \omega_1 , \omega_2 , \omega_3 )$ and similarly for the higher-order blocks.
Recall that according to Eq.~\eqref{eq:g0def} the two-point function for vanishing magnetic field is
$g_0 ( \omega ) = \beta \delta_{\omega,0} b_1$.
Explicit expressions for the relevant generalized blocks are given in Appendix~\ref{app:single_spin}.
The frequency sums can be evaluated analytically and we obtain for vanishing magnetic field
  \begin{equation}
 \Pi^{(2)} ( \bd{k} ,  \omega ) =  \delta_{\omega , 0 }   \frac{\beta^3}{N} \sum_{\bd{q}}
 \left[ 
  \frac{5  b_1 b_3}{6} 
 J^2_{ \bd{q} }  - \frac{b_1^2}{12}  
 J_{ \bd{q} } J_{ \bd{q} + \bd{k} } \right]
 +   ( 1 - \delta_{\omega , 0 } )  \frac{   \beta}{\omega^2}
 \frac{2 b_1^2}{N} \sum_{\bd{q}} J_{ \bd{q} } ( J_{ \bd{q} } - J_{ \bd{q} + \bd{k} } ),
 \end{equation}
where $b_1$ and $b_3$ are defined in Eq.~\eqref{eq:b13def}.
The third-order contribution to the irreducible susceptibility is
represented by the four diagrams in Fig.~\ref{fig:Pi34}. 
The explicit mathematical expression represented by these diagrams
is given in Eq.~\eqref{eq:Pi3int} of Appendix~\ref{app:hightemp}.
After performing the frequency sums we obtain for vanishing magnetic field
 \begin{align}
  \Pi^{(3)} ( \bd{k} , \omega )  = {} & 
- \delta_{\omega,0}  \beta^4 \biggl\{
  \frac{1}{ 72 N^2} \sum_{\bd{q}_1 \bd{q}_2 } \left[
  ( b_1^2 - 10 b_1 b_3 )   J_{ \bd{q}_1 } J_{ \bd{q}_2 } J_{  \bd{q}_1 + \bd{q}_2 }
 +   (  b_1^2 + 20  b_3^2 )  J_{ \bd{q}_1 } J_{ \bd{q}_2 } J_{  \bd{q}_1 + \bd{q}_2 + \bd{k} } \right]
+ \frac{ 5 b_1^2 b_3}{6N} \sum_{\bd{q}} J^3_{\bd{q} }
 \biggr\}
 \nonumber
 \\
 & + ( 1 - \delta_{\omega , 0 } ) \frac{\beta^2}{\omega^2} 
 \biggl\{ \frac{ b_1^2}{ 2 N^2} 
 \sum_{\bd{q}_1 \bd{q}_2 } J_{ \bd{q}_1 } J_{ \bd{q}_2 } \left( 
 J_{  \bd{q}_1 + \bd{q}_2 } - J_{  \bd{q}_1 + \bd{q}_2 + \bd{k} } \right)
 - \frac{ 2 b_1^3}{N} \sum_{\bd{q}} J^2_{ \bd{q} } \left( J_{\bd{q} } - J_{ \bd{q} + \bd{k} } \right)
 \biggr\}.
 \end{align}
\end{widetext}
Note that the exact identity $\Pi ( \bd{k} =0 , \omega \neq 0 ) = 0$ implied by the 
conservation of the total spin is
satisfied order by order in perturbation theory.
It is also straightforward to show that the sum rule
\begin{equation}
\frac{ 1 }{ \beta N } \sum_{ \bm{k} \omega } G ( \bm{k} , \omega ) = b_1 = \frac{ S ( S + 1 ) }{ 3 }
\end{equation}
implied by the spin-length constraint $ \bm{S}_i^2 = S ( S + 1 ) $ is likewise satisfied order by order in perturbation theory.

For a Heisenberg model with nearest-neighbor exchange $J$ on a $D$-dimensional hypercubic lattice 
with lattice spacing $a$ the Fourier transform of the exchange couplings is
$J_{\bd{q}} = J_0 \gamma_{\bd{q} }$ where  $J_0 = 2 D J$ and the 
nearest-neighbor form factor is 
 \begin{equation}
 \gamma_{\bd{q}} = \frac{1}{D} \sum_{\mu =1}^D \cos ( q_{\mu} a ).
 \label{eq:gammaqdef}
 \end{equation}
Using \cite{footnote_nearest_neighbor}
 \begin{subequations} \label{eq:gamma_sums}
 \begin{align}
 \frac{1}{N} \sum_{\bd{q}} \gamma_{\bd{q}} \gamma_{\bd{q} + \bd{k}} & = \frac{\gamma_{\bd{k}}}{2D},
 \\
 \frac{1}{N} \sum_{\bd{q}} \gamma_{\bd{q}}^2 \gamma_{\bd{q} + \bd{k}} & = 0, \label{eq:gamma3_sum}
 \\
 \frac{1}{N^2} \sum_{\bd{q}_1 \bd{q}_2} \gamma_{\bd{q}_1} \gamma_{\bd{q}_2}  \gamma_{\bd{q}_1 + \bd{q}_2 + 
\bd{k}} & = \frac{\gamma_{\bd{k}}}{(2D)^2},
 \end{align}
\end{subequations}
we obtain for the second-order term
 \begin{align}
 \Pi^{(2)} ( \bd{k} , \omega ) = {} & \delta_{\omega , 0 }  \frac{ \beta^3 J_0^2}{ 2 D } \left[ \frac{ 5 b_1 b_3}{6}
 - \frac{ b_1^2 \gamma_{\bd{k}}}{12} \right]
 \nonumber
 \\
 & + ( 1 - \delta_{\omega, 0} ) \frac{ \beta J_0^2}{ 2 D  } 2 b_1^2 
\frac{ 1 - \gamma_{\bd{k}}}{\omega^2} , 
 \label{eq:Pi2res}
 \end{align}
and for the third-order term
 \begin{align}
 \Pi^{(3)} ( \bd{k} , \omega ) = {} & -\delta_{\omega , 0 }  \frac{ \beta^4 J_0^3}{  (2 D)^2  } \left[ 
 \frac{ b_1^2 - 10 b_1 b_3 }{72} + \frac{( b_1^2 + 20 b_3^2) \gamma_{\bd{k}}}{72}  \right]
 \nonumber
 \\
 & + ( 1 - \delta_{\omega, 0} ) \frac{ \beta^2 J_0^3}{ (2 D)^2 }
 \frac{b_1^2}{2} \frac{ 1 - \gamma_{\bd{k}} }{ \omega^2}. 
 \label{eq:Pi3res}
 \end{align}
Comparing our perturbative results at finite frequency $\omega$ with high-temperature expansions for the short-time ($|J|t \ll 1$) evolution of the Kubo relaxation function, we find that our results for $\Pi^{(2)} ( \bd{k} , \omega )$ and $\Pi^{(3)} ( \bd{k} , \omega )$
are consistent with the first two terms in the corresponding second order moment  \cite{Kheli1970,Collins71}. 

\begin{figure}[tb]
 \begin{center}
  \centering
\vspace{7mm}
 \includegraphics[width=0.45\textwidth]{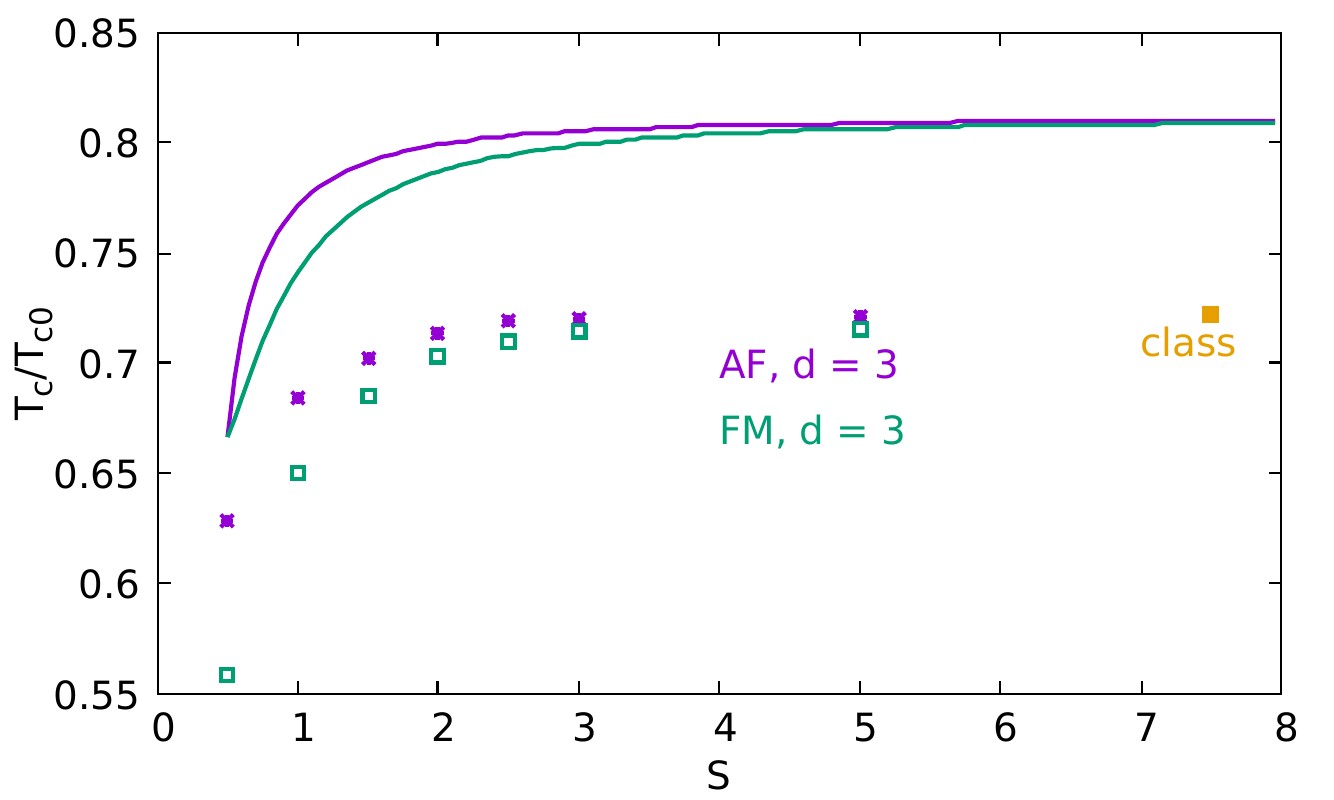}
   \end{center}
  \caption{%
Graphical representation of the results for the critical temperatures of nearest-neighbor ferromagnets or antiferromagnets in a simple cubic lattice listed in Table~\ref{tab:tc}.
The solid lines are obtained from the condition \eqref{eq:Tc2} using the 
order-$J^3$ truncated high-temperature expansion for the inverse static spin susceptibility 
given in Eq.~\eqref{eq:Piinvtrunc}. The crosses and squares represent the benchmarks 
mentioned in the caption of Table~\ref{tab:tc} and additional values for $S \geq 2$ are taken from Ref.~\cite{Cuccoli01}.
}
\label{fig:tc}
\end{figure}

In the static limit $\omega  \rightarrow 0$ we may use the above
results  to estimate the critical temperature $T_c$ where the system develops 
long-range magnetic order. At $T=T_c$ the static susceptibility $ G ( \bd{Q}  , 0 )$ diverges
at the ordering wave-vector $\bd{Q}$. At this temperature
 \begin{equation}
 1 + J_{\bd{Q}} \Pi ( \bd{Q} , 0 ) =0,
 \label{eq:Tc1}
 \end{equation}
or equivalently
 \begin{equation} 
 J_{\bd{Q}} + \Pi^{-1} ( \bd{Q} , 0 ) =0.
 \label{eq:Tc2}
 \end{equation}
If $\Pi ( \bd{Q} , 0 )$ is only known up to some  finite order in $J$, these conditions are not equivalent if we consistently expand  Eqs.~\eqref{eq:Tc1} and \eqref{eq:Tc2} up to this order.
For example, if we expanding the irreducible 
static susceptibility up to third order in $J$,
 \begin{equation}
 \Pi (\bd{k} , 0 ) = \beta b_1 + \Pi^{(2)} ( \bd{k} ,0 ) + \Pi^{(3)} 
( \bd{k} ,0) + {\cal{O}} ( J^4 ),
 \label{eq:Pitrunc}
 \end{equation}
the corresponding expansion of the inverse susceptibility is
 \begin{equation}
 \Pi^{-1} ( \bd{k} , 0 )  = \frac{1}{\beta b_1} - \frac{ \Pi^{(2)} ( \bd{k} ,0 ) }{( \beta b_1)^2} 
- \frac{ \Pi^{(3)} ( \bd{k},0 ) }{( \beta b_1)^2} + {\cal{O}} ( J^4 ). 
 \label{eq:Piinvtrunc}
 \end{equation}
Using this truncated expansion to estimate the
critical temperature $T_c$ in various cases where controlled benchmarks
are available, we obtain the results
summarized in Table \ref{tab:tc} and in Fig.~\ref{fig:tc}.
\begin{table}
\centering
\begin{tabular}{ c c c c c c c c c c}
\hline
\hline
&&&& \multicolumn{3}{c}{$T_c / T_{c0}$} && \multicolumn{2}{c}{rel. error / $\%$} \\
\cline{5-7}
\cline{9-10}
$S$				&&	 $J$	&&	$J^2$	&	$J^3$	&	benchmark	&&	$J^2$	&	$J^3$	\\
\hline
$1/2$ 		&&	$<0$	&&		-	&	0.667	&	0.559	&&		-	&	19.3	\\
$1/2$ 		&&	$>0$	&&	0.667	&	0.667	&	0.629	&&	6	&	6	\\
$1$ 			&&	$<0$	&&	0.645	&	0.741	&	0.650	&&	0.8	&	14  \\
$1$ 			&&	$>0$	&&	0.750	&	0.771	&	0.684		&&	9.6	&	12.7	\\
$3/2$ 		&&	$<0$	&&	0.724	&	0.773	&	0.685		&&	5.7	& 	12.8	\\
$3/2$ 		&&	$>0$	&&	0.769	&	0.791	&	0.702		&&		9.5	& 12.7		\\
$\infty$	&&	$\neq 0$	&&	0.788	&	0.810	&	0.722		&&	9.1	& 	12.2	\\
$\infty (D = 4)$ 	&&	$\neq 0$	&&	0.853	&	0.862	&	0.822		&&	3.7 & 4.9 		
 \\
\hline
\hline
\end{tabular}
\caption{Critical temperatures $T_c$ in units of the mean-field critical temperature $T_{c0} = b_1 | J_{\bd{Q}} |$ 
of spin-$S$ quantum Heisenberg ferromagnets  ($ J < 0$, $\bd{Q} =0$) and
antiferromagnets ($J > 0$, $\bd{Q} = ( \frac{\pi}{a} , \frac{\pi}{a} , \frac{\pi}{a} )$) 
with nearest neighbor exchange $J$. 
With the exception of the last row all temperatures are for a  three-dimensional simple cubic lattice.
The values in the third and fourth columns are obtained by inserting the
high-temperature expansion \eqref{eq:Piinvtrunc} of
$\Pi^{-1} ( \bd{Q} , 0 )$ at order $J^2$ or $J^3$ into Eq.~\eqref{eq:Tc2}.
The benchmarks are obtained using quantum Monte Carlo simulations \cite{Sandvik98,Troyer04} ($S = 1/2$) and a more sophisticated high-temperature series expansion \cite{Oitmaa04}. 
The benchmark in the classical limit ($S= \infty$) in dimension $D=4$  
has been obtained via high-temperature series expansion \cite{McKenzie82}. 
%
}
\label{tab:tc}
\end{table}
It turns out that 
the truncated expansion \eqref{eq:Tc2} of the
inverse susceptibility 
yields more accurate estimates for $T_c$ than Eq.~\eqref{eq:Tc1},
which has already been noticed  by Krieg \cite{Krieg19b}.
In fact, if we substitute the truncated high-temperature expansion \eqref{eq:Pitrunc}
for $\Pi ( \bd{Q} , 0)$ into Eq.~\eqref{eq:Tc1}, we find that
for three-dimensional spin-$S$ Heisenberg ferromagnets or antiferromagnets
with nearest-neighbor coupling the resulting equation does not have any solutions, so that
at this level of approximation we miss the magnetic instability.
On the other hand, if we use the condition \eqref{eq:Tc2} involving $\Pi^{-1} ( \bd{Q} , 0)$
and substitute the corresponding truncated high-temperature expansion 
for $\Pi^{-1} ( \bd{Q} , 0 )$ given in  Eq.~\eqref{eq:Piinvtrunc}
we obtain the expected magnetic instabilities for arbitrary $S$, although
for  a $S=1/2$ ferromagnet the $J^3$-term is crucial to 
reproduce the instability and $T_c$ is roughly $20 \%$ larger than the 
established result. Note also that for $S \geq 1$ 
the $J^2$-truncation produces better estimates for $T_c$ than the 
$J^3$-truncation, 
indicating the alternating nature of the high-temperature series of the
inverse susceptibility. 
A plausible  reason why  
estimates of $T_c$ based on a low-order
truncated expansion of  $\Pi^{-1} ( \bd{Q} , 0)$  produce
better results than the 
corresponding expansion of $\Pi ( \bd{Q} , 0 )$ is that in the former case we expand the physical inverse spin susceptibility in powers of $J / T $, whereas 
$\Pi ( \bd{k}, 0 )$ is an auxiliary quantity that cannot be 
directly measured.
Note also that in Appendix \ref{app:trimer},
we explicitly show for a Heisenberg trimer that a truncated expansion of $\Pi^{-1} ( \bd{k} , 0)$
yields an overall reasonable extrapolation of the static spin susceptibility beyond the perturbative regime,
whereas the truncated expansion of $\Pi ( \bd{k} , 0)$ fails to do so.
Improved estimates for $T_c$ 
can be obtained with moderate 
numerical effort by solving truncated spin-FRG flow equations \cite{Tarasevych22b}. 
In this case it likewise turns out to be crucial to consider the flow of the physical inverse static spin susceptibility $\Pi^{-1} ( \bd{k} , 0)$,
which entails a hybrid approach treating static and dynamic fluctuations differently \cite{Tarasevych21,Tarasevych22b,Rueckriegel24}.
Note finally that our perturbative approach correctly reproduces the
fact that for any finite $S$ the critical temperature  of an antiferromagnet
is always larger than the critical temperature of a ferromagnetic with the same value 
of $| J |$.

Another limit where the critical temperature can be calculated systematically is 
the limit of high dimensions \cite{Krieg19,Krieg19b}.
In fact, by comparing the second-order correction $\Pi^{(2)} ( \bd{k} , \omega )$
in Eq.~\eqref{eq:Pi2res} with the third-order correction 
 $\Pi^{(3)} ( \bd{k} , \omega )$ in Eq.~(\ref{eq:Pi3res}), we see that
in high dimensions successive orders in the high-temperature expansion are additionally controlled by the small parameter $1/(2D)$, as already pointed out in Refs.~[\onlinecite{Krieg19,Krieg19b}].
However, to collect all contributions to $\Pi^{-1} ( \bd{k} , 0 )$ of order
$1/(2D)^2$
we have to take into account contributions
up to order $J^4$ in the high-temperature
expansion
of $\Pi ( \bd{k} , 0 )$ \cite{Krieg19b}.

\subsection{Chiral non-linear susceptibility}

\label{sec:chiral_non-linear_susceptibility}

\begin{figure}[tb]
 \begin{center}
  \centering
\vspace{7mm}
 \includegraphics[width=0.45\textwidth]{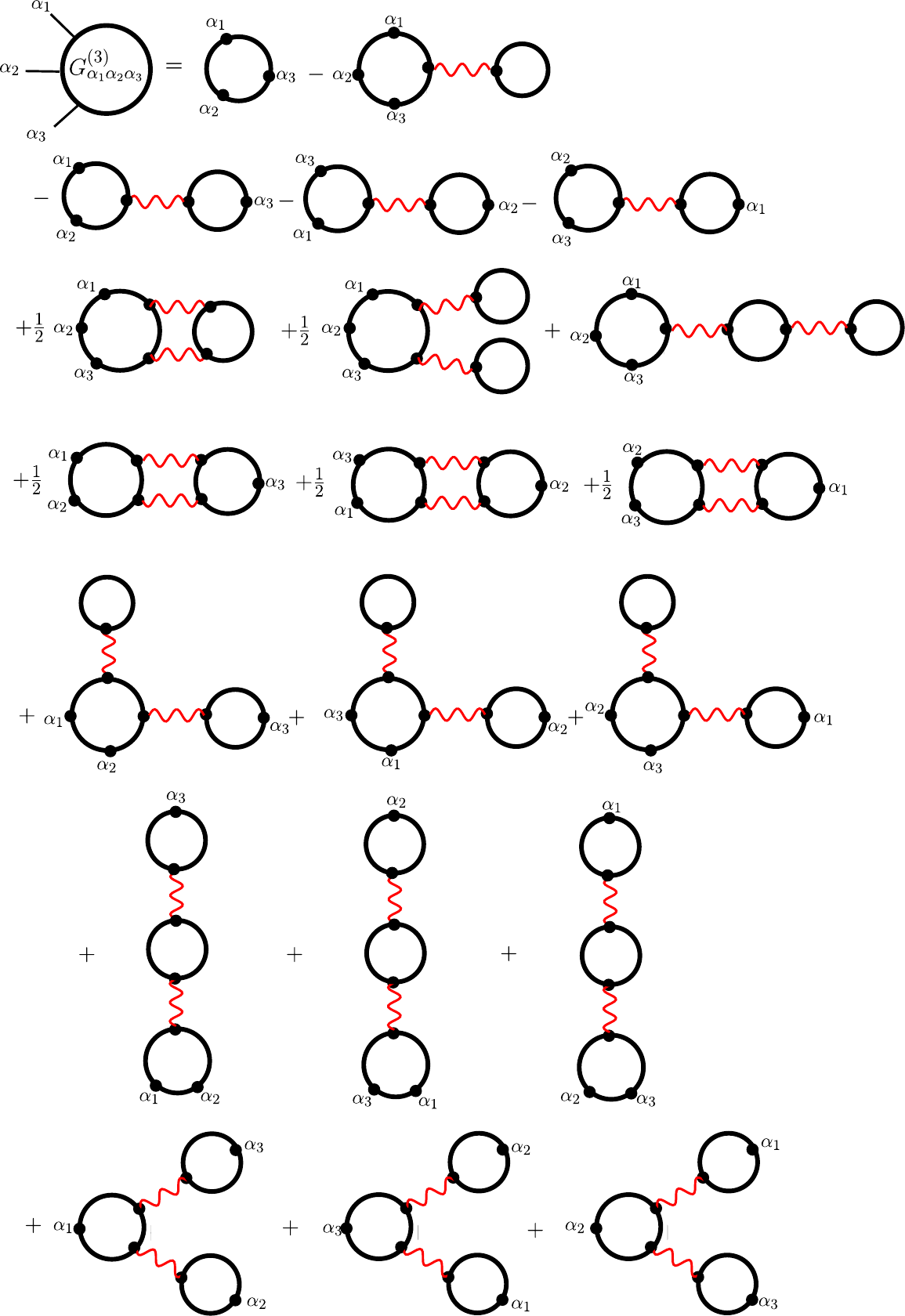}
   \end{center}
  \caption{%
Diagrammatic representation of
the connected three-spin correlation function up to order $J^2$. 
}
\label{fig:G3pert}
\end{figure}

\begin{figure*}[tb]
 \begin{center}
  \centering
\vspace{7mm}
 \includegraphics[width=.95\textwidth]{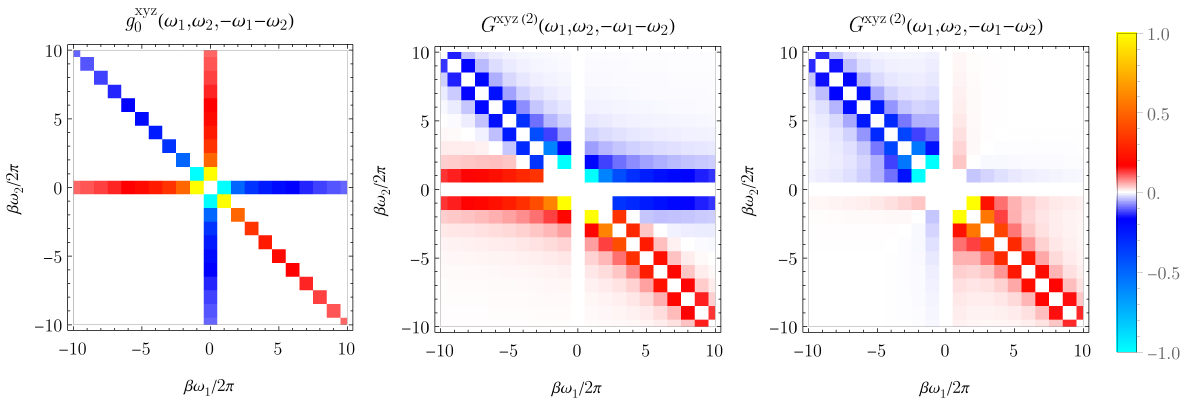}
   \end{center}
  \caption{%
Matsubara-frequency dependence of the chiral quadratic response.
Left: 
Response \eqref{eq:g0chiralCartesian} of an isolated spin,
which is only finite if one of the three frequencies $ \omega_1 $, $ \omega_2 $, or $ \omega_3 = - \omega_1 - \omega_2 $ vanishes. 
Center:
The order-$J^2/T^2$ contribution \eqref{eq:G2chiral} to the chiral 
quadratic response that exhibits the generic frequency dependence,
for $ \lambda_{ \bm{k}_1 } = \lambda_{ \bm{k}_2 } = \lambda_{ \bm{k}_3 } = - 1 $.
Right:
The same for $ \lambda_{ \bm{k}_1 } = \lambda_{ \bm{k}_2 } = 1 $ and $ \lambda_{ \bm{k}_3 } = - 1 $.
All plots are normalized to their maximum value.
}
\label{fig:G3}
\end{figure*}

With our  recursive algorithm it is straightforward to obtain
the high-temperature expansion of arbitrary connected spin correlation functions. 
Of particular interest is the
connected three-spin correlation function, 
which determines the quadratic response of the magnetization
to a time-dependent external magnetic field \cite{Kappl23}. 
In Fig.~\ref{fig:G3pert} we show
all  diagrams contributing to the connected three-spin correlation function up to order $J^2$.
For simplicity let us now 
assume that the spins are located on a Bravais lattice and
are exposed to homogeneous magnetic field $H$ in $z$-direction, so that 
it is again convenient to work in a spherical basis.
In momentum-frequency space the
high-temperature expansion of the chiral three-spin correlation function is 
 \begin{align}
  G^{ +-z } ( K_1 , K_2  , K_3  )    = {} &  g^{+-z } ( \omega_1 , \omega_2 , \omega_3 )
 \nonumber
 \\
  &  +  G^{ +-z (1) } ( K_1 , K_2 , K_3 )
 \nonumber
 \\
 &    +   G^{+-z (2) } ( K_1 , K_2 , K_3 )  +   {\cal{O}} ( J^3 ), 
 \label{eq:Gpmzexp}
 \end{align}
where $K_i = ( \bd{k}_i , \omega_i )$ are multi-labels for momentum and Matsubara frequency
and it is understood that both sides of Eq.~\eqref{eq:Gpmzexp}  should be multiplied by 
 a factor of $\delta_{ \bd{k}_1 + \bd{k}_2 + \bd{k}_3 , 0} \delta_{ \omega_1 + \omega_2 + \omega_3 , 0 }$ implementing conservation of momentum and frequency.
The zeroth-order term  $ g^{+-z } ( \omega_1 , \omega_2 , \omega_3 )$ is given
by the generalized block involving three different spin components
defined in Eq.~\eqref{eq:gchiral}. In the limit $H \rightarrow 0$ 
the chiral three-point function can be written in the
manifestly symmetric form \cite{Tarasevych18,Goll19}, 
 \begin{align}
 g_0^{ +-z}( \omega_1 , \omega_2 , \omega_3 ) = {} &  - \beta b_1
 \left( 1 - \delta_{\omega_1 , 0 } \delta_{\omega_2 , 0 } \delta_{\omega_3 , 0} \right)
 \nonumber
 \\
 & \times
 \left[ \frac{ \delta_{ \omega_1 , 0 } }{ i \omega_2 }
 + \frac{ \delta_{ \omega_2 , 0 } }{ i \omega_3 }
 + \frac{ \delta_{ \omega_3 , 0 } }{ i \omega_1 }
 \right],
 \label{eq:g0chiral}
 \end{align}
where the subscript indicates the zero-field limit and
$b_1 = S ( S+1)/3$ has already been introduced in Eq.~\eqref{eq:b1def}.
Note that for  $H \rightarrow 0$ the purely longitudinal three-point function
$g_0^{zzz} $ vanishes so that in a Cartesian basis
all non-zero  components of the three-point function
can be written
in terms of the three-dimensional Levi-Civita symbol $\epsilon^{a_1 a_2 a_3}$ as 
follows,
 \begin{align}
G^{a_1 a_2 a_3} ( K_1 , K_2 , K_3 )  & =  \epsilon^{a_1 a_2 a_3} 
  G^{xyz} ( K_1 , K_2 , K_3 )
 \nonumber
 \\
 & =  - i \epsilon^{a_1 a_2 a_3} G^{+-z} ( K_1 , K_2 , K_3 ),
 \end{align}
where $\epsilon^{xyz} =1$. In particular, for an isolated spin we obtain
for vanishing magnetic field
\begin{align}
 g_0^{a_1 a_2 a_3}( \omega_1 , \omega_2 , \omega_3 ) = {} &   \epsilon^{a_1 a_2 a_3}
\beta b_1
 \left( 1 - \delta_{\omega_1 , 0 } \delta_{\omega_2 , 0 } \delta_{\omega_3 , 0} \right)
 \nonumber
 \\
 & \times
 \left[ \frac{ \delta_{ \omega_1 , 0 } }{  \omega_2 }
 + \frac{ \delta_{ \omega_2 , 0 } }{  \omega_3 }
 + \frac{ \delta_{ \omega_3 , 0 } }{  \omega_1 }
 \right],
 \label{eq:g0chiralCartesian}
 \end{align}
which is manifestly symmetric under the exchange of any pair of labels
$1 \leftrightarrow 2$, $1 \leftrightarrow 3$ or $ 2 \leftrightarrow 3$.
The frequency dependence of $g_0^{xyz}( \omega_1 , \omega_2 , \omega_3 )$ 
reflects the non-trivial on-site correlations implied by the spin commutation 
relations \cite{Vaks68,Tarasevych18,Goll19}. As a result, 
the  chiral quadratic response is only finite if one of the frequencies in 
 $ g_0^{xyz}( \omega_1 , \omega_2 , \omega_3 )$ vanishes and the other two frequencies are finite.
 This is displayed explicitly in the left panel of Fig~\ref{fig:G3}. 
Obviously, the chiral quadratic response  of an isolated spin does not exhibit 
second-harmonic generation, which is a hallmark of non-linear response.
The fact that the spin dynamics generated by the spin algebra yields 
such a singular
frequency dependence of the chiral quadratic response
has also been pointed out in a recent work by Kappl {\it{et al.}}~\cite{Kappl23}
in the context of the Anderson impurity model.
The frequency dependence of the chiral nonlinear susceptibility of an isolated spin in a magnetic field has recently also been discussed in Ref~[\onlinecite{Halbinger23}].

Let us now explore the fate of this singular 
frequency dependence when we include the effect of the exchange
interaction at high temperatures and frequencies. To first order in $J$ the chiral three-point function
is given by the four first-order diagrams in the first and second lines of Fig.~\ref{fig:G3pert}.
For $H =0$ the tadpole diagram vanishes and the contribution from the three diagrams in the
second line of Fig.~\ref{fig:G3pert} is
 \begin{align}
  & G^{xyz (1)} ( K_1 , K_2 , K_3 )  =   -   \beta^2 b_1^2
 ( 1 - \delta_{\omega_1 , 0} \delta_{\omega_2 , 0}
 \delta_{\omega_3,0} )  
 \nonumber
 \\
 & \times      \left[  J_{\bd{k}_1} \frac{ \delta_{\omega_1,0}}{  \omega_2 }
 +  J_{\bd{k}_2} \frac{ \delta_{\omega_2,0}}{  \omega_3 }  
 + J_{\bd{k}_3} \frac{ \delta_{\omega_3,0}}{ \omega_1 } \right].
  \label{eq:G1chiral}
 \end{align}
At this level of approximation the chiral susceptibility still exhibits the same singular frequency dependence as the generalized block $g_0^{xyz} ( \omega_1 , \omega_2 , \omega_3)$.
To obtain the generic frequency dependence 
of the chiral non-linear susceptibility 
allowing for second-harmonic generation
we have to take into account the  second order in the exchange couplings.
For simplicity let us focus again on the limit $H  \rightarrow 0$ where all tadpole
diagrams involving the one-point functions in Fig.~\ref{fig:G3pert} vanish.
Assuming that  all three Matsubara frequencies $ \omega_1$, $\omega_2$, and
$\omega_3$ are finite,  only four diagrams contribute to the chiral 
three-point function: the diagram involving the five-spin block in the
third row of Fig.~\ref{fig:G3pert}, and the three diagrams where the
four-spin blocks are  connected to the three-spin blocks 
in the fourth row of Fig.~\ref{fig:G3pert}. 
Technical details for the evaluation of these diagrams are given in 
Appendix~\ref{app:hightemp},
where we also give in Eq.~\eqref{eq:Gxyz_complete} the complete second-order high-temperature expansion
of the chiral non-linear susceptibility of an arbitrary spin-$S$ Heisenberg magnet for $ H = 0 $.
If none of the frequencies $ \omega_1$, $\omega_2$, and $\omega_3$ are zero
we obtain for vanishing magnetic field
 \begin{align}
  G^{xyz (2)} ( K_1 , K_2 , K_3 ) = {} & 
  \beta b_1^2 \left( \frac{1}{N} \sum_{\bd{q}} J_{\bd{q} }^2 \right)
   \nonumber
   \\
   & \times  \Biggl[ \frac{ 1 - \lambda_{\bd{k}_1} }{\omega_1^2}
   \left( \frac{1}{\omega_2} -  \frac{1}{\omega_3} \right)
   \nonumber
   \\ 
    & \hspace{5mm} + \frac{ 1 - \lambda_{\bd{k}_2} }{\omega_2^2}
   \left( \frac{1}{\omega_3} -  \frac{1}{\omega_1} \right)
   \nonumber
    \\ 
    & \hspace{5mm} + \frac{ 1 - \lambda_{\bd{k}_3} }{\omega_3^2}
   \left( \frac{1}{\omega_1} -  \frac{1}{\omega_2} \right)
   \Biggr],
    \label{eq:G2chiral}
 \end{align}
where the momentum-dependent form factor $\lambda_{\bd{k}}$
is defined by the following ratio of Brillouin zone averages: 
 \begin{equation}
 \lambda_{\bd{k}} = \frac{ \sum_{\bd{q}} J_{\bd{q}} J_{\bd{q} + \bd{k}} }{
  \sum_{\bd{q}} J^2_{\bd{q}}}.
 \label{eq:fkdef}
  \end{equation}
For nearest-neighbor exchange the form factor 
$\lambda_{\bd{k}} = \gamma_{\bd{k}}$ 
agrees with nearest-neighbor form factor defined in Eq.~\eqref{eq:gammaqdef}.
For the special case $S=1/2$  the 
frequency-structure in Eq.~\eqref{eq:G2chiral} has been obtained previously 
by Krieg \cite{Krieg19b}, who has calculated the corresponding
irreducible three-point vertex.
This frequency dependence is shown in the center and right panels of Fig.~\ref{fig:G3} for two different sets of wave vectors. 
Note that the single-spin result \eqref{eq:g0chiralCartesian} and the
first-order correction \eqref{eq:G1chiral} to the chiral non-linear susceptibility are only finite
if one of the frequencies $\omega_1$, $\omega_2$, or $\omega_3$ vanishes and the
other two frequencies are non-zero; the corresponding quadratic response 
does not exhibit second-harmonic generation.
In contrast, the
second-order non-linear chiral spin  susceptibility \eqref{eq:G2chiral} is finite for
all frequencies compatible with frequency conservation, as illustrated in 
Fig.~\ref{fig:G3}.
In particular, for $\omega_1 = \omega_2 = \omega \neq 0$ and $\omega_3 = - 2 \omega $ we obtain
 \begin{align}
  & G^{xyz (2)} ( \bd{k}_1 , \omega , \bd{k}_2 ,  \omega, - \bd{k}_1 - \bd{k}_2 , - 2 \omega )
  \nonumber
   \\
   & =
    - 3  \beta b_1^2 \left( \frac{1}{N} \sum_{\bd{q}} J_{\bd{q} }^2 \right)
     \frac{ \lambda_{\bd{k}_1} - \lambda_{\bd{k}_2}}{ 2 \omega^3}.
  \end{align}
Note that the frequencies in the above expressions are bosonic Matsubara frequencies so that an analytic continuation to the real frequency axis is necessary to obtain the physical
chiral response function.
We conclude that a necessary condition for
observing second-harmonic generation in the 
chiral non-linear response function of a quantum spin system 
at high temperatures and frequencies is
$\lambda_{\bd{k}_1} \neq \lambda_{\bd{k}_2 }$, which means that the
external magnetic field must be inhomogeneous such that the spatial dependence of
two orthogonal components of the field is characterized by  
different wave-vectors.

\section{Summary and conclusions}

In this work we have derived  a new recursive algorithm for generating  the
high-temperature series expansion of the free energy and arbitrary connected correlation functions of quantum spin systems. Our algorithm is based on a formally exact 
functional renormalization group flow equation for the generating functional of imaginary-time-ordered connected spin correlation functions \cite{Krieg19}. Using a specific interaction-switch deformation scheme, we have obtained a system of recursion relations given by Eq.~\eqref{eq:iterate}
which can be iteratively solved to obtain the coefficients in the high-temperature expansion of arbitrary  connected correlation functions. 
It is easy to implement our iterative algorithm  
using symbolic manipulation software such as  MATHEMATICA to obtain the free energy, the magnetic equation of state, and the two-point function up to order $J^4$. 
The main advantages of our method are as follows:
\begin{enumerate}
\item
Our algorithm allows us to systematically compute dynamical multi-spin correlation functions
of spin-$S$ quantum magnets in a unified framework.
No additional combinatorial considerations are necessary;
all combinatorial factors are  explicitly taken into account via
the symmetrization operator 
 ${\cal{S}}_{ 1, \ldots, m; m+1, \ldots, n }$ defined in
Eq.~\eqref{eq:symdef}.
\item
Our formulas are fully analytical.
This is particularly important for $n$-spin functions with $ n > 2 $,
since storing them numerically for all wave-vectors and Matsubara frequencies requires huge data sets for larger $n$. 
It also facilitates the
eventual analytical continuation of dynamical correlation functions to real frequencies.
\item
Our algorithm yields high-temperature series that are valid for arbitrary lattices as well as arbitrary spin $S$,
not just $ S = 1 / 2 $, 
without any additional effort.
\end{enumerate}

Whether or not a fully numerical implementation of our recursive algorithm
based on Eq.~\eqref{eq:iterate} can be used to obtain competitive high-order results
in the high-temperature series expansion remains an interesting open question. 
For Heisenberg models, one has to invest some effort into calculating 
the initial conditions for the iteration given by the generalized blocks describing the
frequency-dependent correlations between different components of a single spin.
These correlations can be calculated iteratively using the recursive form
of the generalized Wick theorem for spin operators \cite{Goll19,Halbinger23}.
On the other hand, for Ising models the initial conditions can be explicitly written down in 
closed form. Note that the
spin-$S$ Ising model can be obtained from our general spin Hamiltonian \eqref{eq:Hamiltonian}
by setting $J^{ab}_{ij} = \delta^{az} \delta^{bz}J_{ij} $.
In this case all operators in our spin Hamiltonian commute so that their 
correlation functions are time-independent. For vanishing exchange couplings the connected $n$-point functions are then given by \cite{Goll19}
 \begin{equation}
  G^{(n,0)}_{ i_1 \ldots i_n} = \delta_{ i_1 i_2 } \delta_{i_2 i_3}
   \ldots \delta_{i_{ n-1} i_n } b^{(n-1)} ( \beta H ),
   \end{equation}
where $b^{(n-1)} ( y)$ is the $(n-1)$-st derivative of the 
spin-$S$ Brillouin function $b(y)$ defined in Eq.~\eqref{eq:bdef}.
Given this initial condition, a fully numerical implementation of our
recursive algorithm for Ising models seems to be straightforward.

We have used our method to calculate the interaction-irreducible spin susceptibility $\Pi ( \bd{k} , \omega )$ to order $J^3$ and the chiral non-linear spin susceptibility 
$G^{xyz} ( K_1 , K_2 , K_3)$ to order $J^2$. From our perturbative result for the
 static part  $\Pi^{-1} ( \bd{k} , 0 )$
 of the {\it{inverse}} interaction-irreducible susceptibility we have 
obtained 
estimates for the critical temperature of three-dimensional nearest-neighbor Heisenberg ferro- and antiferromagnets
which overestimate the known results by roughly $10$-$20 \%$. 
Given the simplicity of our method it should be useful to estimate
critical temperatures of more complicated spin systems where accurate benchmarks 
are not readily available \cite{Rau14,Smejkal23,Gaitan24}.

The fact that the chiral non-linear susceptibility in strongly correlated systems 
exhibits an interesting frequency dependence has recently 
been pointed out by Kappl {\it{et al.}}~\cite{Kappl23} who presented numerical results for the
chiral non-linear susceptibility of the Anderson impurity model.
Surprisingly, a thorough investigation of this quantity  for  
Heisenberg magnets at low temperatures seems not to exist in the literature.
Although the perturbative result for the chiral non-linear susceptibility
in Eq.~\eqref{eq:G2chiral} is only valid at high temperatures and
frequencies, it exhibits a rather non-trivial momentum- and frequency dependence which can
in principle be observed  experimentally by measuring
the strength of second-harmonic
generation in the chiral quadratic magnetization response 
of Heisenberg magnets.
Moreover,
the comparison to the exact solution of the Heisenberg trimer in Appendix \ref{app:trimer}
suggests that this non-trivial momentum-frequency dependence persists also beyond the high-temperature regime.

Finally, it should be mentioned that for classical spin models the idea 
to generate high-temperature expansions from FRG flow equations has been put forward previously 
by Jacquin and Ran{\c c}on \cite{Jacquin16}. They
used the  Wetterich equation \cite{Wetterich93} (i.e., the exact flow equation for the generating functional of irreducible vertices) to
expand the free energy of a classical spin model up to fourth order in $\beta J$.
In contrast, our approach is based on the flow equation for the generating functional of connected correlation functions; this has the advantage of leading to an explicit
recursion given in Eq.~(\ref{eq:iterate}) which can be iterated to obtain high-temperature expansions of arbitrary time-ordered correlation functions of classical and quantum spin systems.

{\it{Note added:}}
After submission of this work a preprint
by Schneider {\it{et al.}} \cite{Schneider24}
has appeared where the spin diagram technique has been used to obtain  high-temperature expansions of the two-point function of Ising and Heisenberg models up to fourth order in the exchange couplings. 
Schneider {\it{et al.}} \cite{Schneider24}  obtained
the relevant diagrams following the perturbative strategy pioneered
by Vaks, Larkin, and Pikin \cite{Vaks68,Vaks68b}. 
In contrast, in this work we have used  FRG flow 
equations to generate the perturbation series.  In  cases where the perturbative expressions presented here
can be compared with those given in Ref.~[\onlinecite{Schneider24}] the results agree.

\begin{acknowledgments}
%
%
We thank Bj\"{o}rn Sbierski for useful discussions.
This work was financially supported by the Deutsche
Forschungsgemeinschaft (DFG, German Research Foundation) 
through Project No. 431190042.
\end{acknowledgments}

\appendix

\renewcommand{\appendixname}{APPENDIX}

\renewcommand{\thesection}{\Alph{section}}

\section{Connected correlations of a single spin in a magnetic field}

\label{app:single_spin}

\renewcommand{\theequation}{A\arabic{equation}}

\setcounter{equation}{0}
\setcounter{subsection}{0}

In this appendix we summarize the connected imaginary-time-ordered correlation functions
for the components of a single spin-$S$  operator $\bd{S}$ in a magnetic field. 
Following Ref.~[\onlinecite{Izyumov88}], 
we call these quantities {\it{generalized blocks}}. 
Choosing our coordinate system such that the $z$-axis is aligned with the direction of the magnetic field, the Hamiltonian is simply $ {\cal{H}}_0 =  - H S^z$, where 
the magnetic field  $H$ is measured in units of energy. 
Although for  $H \rightarrow 0$ the Hamiltonian vanishes, the spin correlations remain non-trivial
due to the SU(2)-commutation relations between different spin components.
The generalized blocks involving up to four spin components
can be found in Refs.~[\onlinecite{Vaks68,Vaks68b,Izyumov88,Goll19}].
For the calculation of the interaction-irreducible spin susceptibility $\Pi ( \bd{k} , \omega )$
to order $J^3$ and the chiral non-linear 
susceptibility $G^{xyz} ( K_1 , K_2 , K_3 )$ to order $J^2$
in Sec.~\ref{sec:trunc} we also need 
the generalized five-spin blocks which can be 
obtained from the 
recursive algorithm described in the Appendix B of Ref.~[\onlinecite{Goll19}]. A similar  algorithm has recently been constructed in Ref.~[\onlinecite{Halbinger23}].

For finite magnetic field it is convenient to work with the spherical spin components
$S^p$ defined in Eq.~\eqref{eq:spherical}, where $p=+,-$ labels the two circular components transverse to the magnetic field and $p = 0 = z$ 
labels the longitudinal component 
in the direction of the field.
In this basis the connected two-point functions are
given in Eqs.~\eqref{eq:gzz} and \eqref{eq:gpp}.
The connected three-point functions are more interesting. 
Of particular interest is the chiral 
three-point function $g^{+-z} ( \omega_1 , \omega_2 , \omega_3 )$  
involving three different spin components which can be written as~\cite{Goll19}
 \begin{equation}
 b g^{+-z} ( \omega_1 , \omega_2 , \omega_3 ) = - g ( \omega_1 ) \left[
 g ( - \omega_2 ) -  \beta \delta_{\omega_3,0} b^{\prime} \right].
 \label{eq:gchiral}
\end{equation} 
Here
 \begin{equation}
 g ( \omega ) = \frac{ b}{H - i \omega },
 \end{equation}
where  $ b = b( \beta H )$ is the Brillouin function 
and $b^{\prime} = b^{\prime} ( \beta H )$
is its  first derivative. 
For finite magnetic field $H$ the purely longitudinal three-point function is also finite,
 \begin{equation}
 g^{zzz} ( \omega_1 , \omega_2 , \omega_3 ) = \beta^2 \delta_{\omega_1 , 0 }
 \delta_{\omega_2, 0} b^{\prime \prime} ( \beta H ),
 \end{equation}
where
$b^{\prime \prime} ( \beta H  )$ 
is the second derivative of the Brillouin function.
Recall that with our convention \eqref{eq:frequencycon} 
the above three-point functions
should be multiplied by an overall
factor of $\beta \delta_{ \omega_1  + \omega_2 + \omega_3 , 0 }$ so that
we can use this constraint to write our correlation functions in a more symmetric form.
For  $H \rightarrow 0$  the longitudinal three-point function vanishes because $b^{\prime \prime} (0) =0$  
while the
chiral three-point function \eqref{eq:gchiral} has a non-trivial limit
which can be written in the highly symmetric form given 
in Eq.~\eqref{eq:g0chiral}. 

Next, consider the connected four-point functions in a magnetic field.
The purely longitudinal four-point function is
 \begin{equation}
 g^{zzzz} ( \omega_1 , \omega_2 , \omega_3,\omega_4 ) = \beta^3 \delta_{\omega_1 , 0 }
 \delta_{ \omega_2, 0} \delta_{\omega_3,0} b^{\prime \prime \prime} ( \beta H ),
 \end{equation}
where for $H \rightarrow 0$ the third derivative $b^{\prime \prime \prime} ( \beta H) $ 
of the Brillouin function has a finite limit $b_3$ given by Eq.~(\ref{eq:b3def}).
The four-point functions involving transverse spin components are more interesting.
The purely transverse four-point function can be written as \cite{Goll19}
 \begin{align}
 & b^2 g^{++--} ( \omega_1 , \omega_2 , \omega_3 , \omega_4 ) =  - g ( \omega_1 )   g ( \omega_2 ) \Bigl\{
 \nonumber
 \\
 &       g ( - \omega_3 ) +   g ( - \omega_4 )  -
 \beta \left[  \delta_{  \omega_1 , -  \omega_3} + \delta_{ \omega_1 , - \omega_4} \right]  b^{\prime} \Bigr\},
 \label{eq:Gppmm}
 \end{align}  
while the mixed transverse-longitudinal four-point function is
 \begin{align}
 &   b^2 g^{+-zz} ( \omega_1 , \omega_2 , \omega_3 , \omega_4 ) =
 g ( \omega_1 )   g ( - \omega_2 ) \Bigl\{
 \nonumber
 \\
 & 
  g ( \omega_1 +  \omega_3 ) +   g ( \omega_1 +  \omega_4 )  - 
 \beta \left[  \delta_{  \omega_3,0} + \delta_{ \omega_4,0} \right]  b^{\prime}  \Bigr\}
 \nonumber
 \\
 & + g ( \omega_1 ) \beta^2 \delta_{\omega_{3}, 0} \delta_{\omega_4,0}  b b^{\prime \prime} .
 \label{eq:Gpmzz}
 \end{align}  
For vanishing magnetic field only special frequency combinations
of Eqs.~\eqref{eq:Gppmm} and \eqref{eq:Gpmzz} have a non-zero limit.
Here we give only those contributions needed for the
evaluation of the second-order contribution  $\Pi^{(2)} ( \bd{k} , \omega )$
to the irreducible susceptibility for vanishing magnetic field given
in Eq.~\eqref{eq:Pi2int},
 \begin{align}
 g_0^{zzzz} (0,0, \omega , - \omega ) & = \beta^3 \delta_{\omega , 0 } b_3,
 \\
 g_0^{ +-zz} ( 0,0, \omega , - \omega ) & =  \frac{\beta^3}{3}  
 \delta_{\omega , 0 }  b_3 +  2 \beta (1 - 
 \delta_{\omega , 0 } ) \frac{  b_1}{ \omega^2 }.
 \label{eq:g4zero}
 \end{align}

For the calculation of the third-order correction   $\Pi^{(3)} ( \bd{k} , \omega )$
to the irreducible susceptibility
we also need the generalized five-spin blocks
which can be obtained either diagrammatically following Vaks, Larkin, and Pikin \cite{Vaks68}, or purely algebraically using the 
recursive form 
the generalized Wick theorem for spin operators derived in Appendix B of Ref.~[\onlinecite{Goll19}]. We obtain
 \begin{widetext}
 \begin{align}
 & b^3 g^{++--z} ( \omega_1 , \omega_2 , \omega_3 , \omega_4 , \omega_5 )  =  
 g ( - \omega_3 ) g ( - \omega_4 ) \biggl\{
 \nonumber
 \\
 &
 g ( \omega_1 ) \Bigl[  g ( \omega_1 + \omega_5 ) + g ( - \omega_3 - \omega_5 ) 
 + g ( - \omega_4 - \omega_5 ) 
 - \beta [ \delta_{\omega_5,0} + \delta_{\omega_2, - \omega_3} + \delta_{\omega_2, - \omega_4} ] b^{\prime} 
\Bigr]
 \nonumber
 \\
 + \, &   g ( \omega_2 ) \Bigl[  g ( \omega_2 + \omega_5 ) + g ( - \omega_3 - \omega_5 ) 
 + g ( - \omega_4 - \omega_5 ) 
 - \beta [ \delta_{\omega_5,0} + \delta_{\omega_1, - \omega_3} + \delta_{\omega_1, - \omega_4} ] b^{\prime} 
\Bigr] 
 \nonumber
 \\
 + \, & \beta^2 \delta_{\omega_5,0} [ \delta_{ \omega_1 , - \omega_3 } + \delta_{\omega_2, - \omega_3 } ] b b^{\prime \prime}
\bigg\},
 \label{eq:G5a}
 \end{align}
and for the other mixed five-spin block,
 \begin{align}
 &   b^3 g^{+-zzz} ( \omega_1 , \omega_2 , \omega_3 , \omega_4 , \omega_5 )    =  
 - g ( \omega_1 ) g ( - \omega_2 ) \biggl\{
 \nonumber
 \\
 & 
 \Bigl[ g ( \omega_1 + \omega_3 ) - \beta \delta_{\omega_3,0} b^{\prime} \Bigr]
 \Bigl[ g ( - \omega_2 - \omega_4 ) + g ( - \omega_2 - \omega_5 ) \Bigr]
 + \beta^2 \delta_{\omega_4,0} \delta_{\omega_5,0} b b^{\prime \prime}
 \nonumber
 \\
 + \, & 
  \Bigl[ g ( \omega_1 + \omega_4 ) - \beta \delta_{\omega_4,0} b^{\prime} \Bigr]
 \Bigl[ g ( - \omega_2 - \omega_5 ) + g ( - \omega_2 - \omega_3 ) \Bigr]
 + \beta^2 \delta_{\omega_5,0} \delta_{\omega_3,0} b b^{\prime \prime}
 \nonumber
 \\
 + \,  & 
 \Bigl[ g ( \omega_1 + \omega_5 ) - \beta \delta_{\omega_5,0} b^{\prime} \Bigr]
 \Bigl[ g ( - \omega_2 - \omega_3 ) + g ( - \omega_2 - \omega_4 ) \Bigr]
  + \beta^2 \delta_{\omega_3,0} \delta_{\omega_4,0} b b^{\prime \prime}
 \bigg\}
 \nonumber
 \\
 + \, &  g ( - \omega_2 ) \beta^3 \delta_{\omega_3,0} \delta_{\omega_4,0} \delta_{\omega_5,0} b^2 b^{\prime \prime \prime}.
 \label{eq:G5b}
 \end{align}

\end{widetext}

\section{Evaluation  
of high-temperature expansions}

\label{app:hightemp}

\renewcommand{\theequation}{B\arabic{equation}}

\setcounter{equation}{0}
\setcounter{subsection}{0}

In this appendix, we give some technical details for the evaluation of the 
perturbative high-temperature expansions
in Sec.~\ref{sec:trunc}.  Consider first the high-temperature expansion
of the negative free energy $- \beta F = G^{(0)}$ in units of temperature given in 
Eq.~(\ref{eq:expansionfree}). The third-order coefficient $G^{(0,3)}$ represented by the five
third-order diagrams in Fig.~\ref{fig:freeenergy} is given by
\begin{widetext}
 \begin{align}
G^{(0,3)} = {} & - \frac{1}{12 \beta} \sum_{ij} 
 \sum_{ a_1 b_1} \sum_{a_2 b_2 } \sum_{ a_3 b_3}
 J_{ij}^{ a_1 b_1} J_{ij}^{ a_2 b_2}
 J_{ij}^{a_3 b_3} 
 \sum_{ \omega \omega^{\prime}}
 g^{ a_1 a_2 a_3}_{i} ( \omega , \omega^{\prime} , - \omega - \omega^{\prime} )
 g^{ b_1 b_2 b_3}_{j} ( - \omega , - \omega^{\prime} ,  \omega  +  \omega^{\prime} )
 \nonumber
 \\
 & - \frac{1}{6} \sum_{ijk} \sum_{ a_1 b_1} \sum_{ a_2 b_2 } \sum_{ a_3 b_3} J^{a_1 b_1}_{ij} J^{ a_2 b_2}_{ jk} J^{a_3 b_3}_{ki}
 \sum_{\omega} g_i^{ a_1 b_3} ( \omega ) g_j^{ a_2 b_1} ( \omega )
 g_k^{a_3 b_2} ( \omega )
 \nonumber
 \\
 & - \frac{\beta}{6}  \sum_{ijkl} \sum_{ a bc}
 J^{ a z }_{ij} J^{  b z}_{ i k} 
J^{ c z }_{il}  g^{ abc}_i ( 0,0,0)  m_j m_k m_l 
 \nonumber
 \\
 & - \frac{1}{2} \sum_{ ijk} \sum_{ a_1 b_1} \sum_{ a_2 b_2 } \sum_c
 J_{ij}^{ a_1 b_1}  J_{ ij}^{ a_2 b_2} J_{ ik}^{ c z} 
 \sum_{\omega} g_i^{ a_1 a_2 c} ( \omega , - \omega , 0 ) g_j^{ b_2 b_1} ( \omega ) m_k
 \nonumber
 \\
& - \frac{\beta}{2} \sum_{ ijkl} \sum_{ a b cd}
 J^{ a b}_{ ij} J_{ ik}^{c z} J_{ jl}^{d z}
 g_i^{a c } (0) g_j^{b d} (0)  m_k m_l .
 \label{eq:G03}
  \end{align}
For the special case of an isotropic Heisenberg magnet ($J^{ab}_{ij} = \delta^{ab} J_{ij})$ 
in a homogeneous magnetic field (${H}_i = H$) along the $z$-axis the generalized blocks 
$g^{ a_1 \ldots a_n} ( \omega_1 , \ldots , \omega_n )$
are independent of the site label. The resulting high-temperature expansion of the free energy 
in a magnetic field is then
 \begin{align}
 F = {} &  - \frac{ N B ( \beta H ) }{\beta} + \frac{m^2}{2} \sum_{ ij} J_{ij} 
  -  \frac{1}{4  \beta } \sum_{ ij} J^2_{ ij}  \sum_{\omega}   \sum_{a b} g^{ab} ( \omega ) g^{ba} (  \omega )  
 - \frac{m^2}{2} g^{zz} (0) \sum_{ ijk} J_{ij} J_{jk}
 \nonumber
 \\
 &  + \frac{1}{12 \beta^2 } \sum_{ij} J_{ij}^3\sum_{ \omega \omega^{\prime}} \sum_{abc}
 g^{abc} ( \omega , \omega^{\prime} , - \omega - \omega^{\prime} )
 g^{abc} ( - \omega , - \omega^{\prime} ,  \omega  + \omega^{\prime} )
 \nonumber
 \\
 & + \frac{1}{6 \beta } \sum_{ijk} J_{ij} J_{jk} J_{ki} \sum_{\omega}  \sum_{abc}
 g^{ab} ( \omega ) g^{bc} ( \omega ) g^{ca} ( \omega ) 
 + \frac{ m^3}{6} g^{zzz} (0,0,0) \sum_{ijkl} J_{ij} J_{ ik} J_{il}
 \nonumber
 \\
 & + \frac{ m}{2 \beta} \sum_{ijk} J^2_{ij} J_{ ik} \sum_{\omega}  \sum_{ab} g^{abz} ( \omega , - \omega , 0 ) g^{ba} ( \omega )
+
\frac{ m^2}{2} \sum_{ijkl} J_{ij} J_{ik} J_{ jl} \sum_a [ g^{az} (0) ]^2
 + {\cal{O}} ( J^4).
 \end{align}
The evaluation of the frequency sums is straightforward.
In the zero-field limit $ H \to 0 $,
this yields the following high-temperature expansion
for the free energy of a spin-$S$ Heisenberg magnet:
\begin{align}
F = {} & - \frac{ N }{ \beta } \ln \left( 2 S + 1 \right)
- \frac{ 3 }{ 4 } \beta b_1^2 \sum_{ i j } J_{ i j }^2
- \frac{ 1 }{ 8 } \beta^2 b_1^2 \sum_{ i j } J_{ i j }^3
+ \frac{ 1 }{ 2 } \beta^2 b_1^3 \sum_{ i j k } 
J_{ i j } J_{ j k } J_{ k i }
+ {\cal{O}} ( J^4) .
\label{eq:F_third_order}
\end{align}

Next, consider the high-temperature expansion of the two-spin correlation function,
 \begin{equation}
 G^{ ab }_{ij} ( \omega )   = \delta_{ij}  g_i^{ab} ( \omega ) + 
 G^{ ab (1) }_{ij} ( \omega ) +  G^{ ab (2) }_{ij} ( \omega ) +
 G^{ ab (3)}_{ij} ( \omega )  + {\cal{O}} ( J^4 ),
 \end{equation}
where the zeroth-order term $g_i^{ab} ( \omega )$ is given in
Eq.~\eqref{eq:gabCartesian}, the first-order term 
$G^{ab(1)}_{ij} ( \omega )$ is given 
in Eq.~\eqref{eq:Gab1}, and the second-order term is
 \begin{align}
 G^{ ab (2) }_{ij} ( \omega ) = {} &
 \frac{1}{2 \beta} \sum_{ a_1 b_1} \sum_{ a_2 b_2} J^{ a_1 b_1}_{ij} J^{a_2 b_2}_{ij}
  \sum_{\omega^{\prime}} g^{a a_1 a_2}_{i} ( \omega , \omega^{\prime} , - \omega - \omega^{\prime} )
 g^{ b b_1 b_2}_j ( - \omega , - \omega^{\prime} , \omega + \omega^{\prime} )
 \nonumber
 \\
 & +   \frac{\delta_{ij}}{2 \beta} \sum_k 
\sum_{ a_1 b_1} \sum_{ a_2 b_2} J^{ a_1 b_1}_{ik} J^{a_2 b_2}_{ik}
 \sum_{\omega^{\prime}} g^{a b a_1 a_2}_{i} ( \omega , - \omega , 
\omega^{\prime} , - \omega^{\prime} )
 g^{ b_1 b_2}_k ( \omega^{\prime} , - \omega^{\prime} ) 
 \nonumber
 \\
 & + \sum_k \sum_{ a_1 b_1} \sum_{ a_2 b_2} J^{ a_1 b_1}_{ik} J^{a_2 b_2}_{kj}
 g^{a a_1}_i ( \omega ) g^{ b_1 a_2}_k ( \omega ) g^{b_2 b}_j ( \omega )
 \nonumber
 \\
 & + \frac{\delta_{ij}}{2} \sum_{kl} \sum_{cd}  J^{c z}_{ik} J^{d z}_{il}
 g^{ab cd}_{i} ( \omega , - \omega , 0,0) m_k m_l 
 \nonumber
 \\ 
 & + \delta_{ij} \sum_{kl} \sum_{ a_1 b_1 c}
 J^{a_1 b_1}_{ ik} J^{ cz}_{ kl} g^{aba_1}_i ( \omega , - \omega , 0 )
 g^{b_1 c}_{k} (0) m_l
 \nonumber
 \\
 & + \sum_k \sum_{a_1 b_1 c}  J_{ij}^{a_1 b_1} \left[ J_{ik}^{cz} 
 g^{aa_1 c}_{i} ( \omega , - \omega , 0 ) g^{b_1 b}_j ( \omega )
 + J_{jk}^{cz} g^{ a a_1}_i (\omega) g^{b_1 b c}_j ( \omega , - \omega , 0 ) \right]
 m_k. 
 \label{eq:G22}
 \end{align}
Using Eq.~\eqref{eq:g4zero} we can easily evaluate
the frequency sums in our expression for 
$\Pi^{(2)} ( \bd{k} , \omega )$ in Eq.~\eqref{eq:Pi2int},
 \begin{align}
 & \frac{1}{ 2 \beta }  
  \sum_{ \omega^{\prime}} g_0 (  \omega^{\prime} )
  \Bigl[ 
   2 g_0^{+-zz} ( \omega^{\prime} , - \omega^{\prime} , \omega , - \omega ) 
  + g_0^{zzzz} ( \omega^{\prime} , - \omega^{\prime} , \omega , - \omega )  \Bigr] =
 \delta_{\omega , 0} \beta^3 \frac{5}{6} b_1 b_3 + ( 1 - \delta_{\omega,0} )
 \frac{ \beta}{\omega^2} 2 b_1^2,
 \\
 & \frac{1}{\beta} 
  \sum_{ \omega^{\prime}} 
 g_0^{+-z} ( \omega^{\prime} , - \omega - \omega^{\prime} , \omega )
 g_0^{+-z} ( \omega + \omega^{\prime} ,  - \omega^{\prime} , - \omega )
 =  - \delta_{\omega , 0} \beta^3 \frac{b_1^2}{12}  - ( 1 - \delta_{\omega,0} )
 \frac{ \beta}{\omega^2} 2 b_1^2.
 \end{align}
The third-order correction  $\Pi^{(3)} ( \bd{k} ,  \omega )$ to the irreducible spin susceptibility represented by the four diagrams in Fig.~\ref{fig:Pi34} is
 \begin{align}
\Pi^{(3)} ( \bd{k} , \omega )  = {} &  
- \frac{1}{N^2 \beta^2}
 \sum_{\bd{q}_1 \bd{q}_2} J_{ \bd{q}_1 } J_{ \bd{q}_2 } J_{  \bd{q}_1 + \bd{q}_2 }
 \sum_{\omega_1 \omega_2}
 g^{+-zzz} ( \omega_1 , \omega_2 , - \omega_1 - \omega_2 , \omega , - \omega )
 g^{+-z} ( - \omega_2 , - \omega_1 , \omega_1 + \omega_2 )
 \nonumber
 \\
 & -     \frac{1}{N^2 \beta^2}
 \sum_{\bd{q}_1 \bd{q}_2} J_{ \bd{q}_1 } J_{ \bd{q}_2 } J_{  \bd{q}_1 + \bd{q}_2 + \bd{k} }
 \sum_{\omega_1 \omega_2}
 \Bigl[
 g^{+-zz} ( \omega_1 , \omega_2 , \omega , - \omega_1 - \omega_2 - \omega)
 g^{+-zz} ( - \omega_2 , - \omega_1 , - \omega, \omega_1 + \omega_2 + \omega ) 
 \nonumber
 \\
 &  \hspace{48mm} + \frac{1}{6}  g^{zzzz} ( \omega_1 , \omega_2 , \omega , - \omega_1 - \omega_2 - \omega)
 g^{zzzz} ( - \omega_2 , - \omega_1 , - \omega, \omega_1 + \omega_2 + \omega ) 
 \Bigr]
 \nonumber
 \\
 & -  \frac{1}{2 \beta N} \sum_{\bd{q}} J^3_{\bd{q} } \sum_{\omega^{\prime}}
 g^2 ( \omega^{\prime} ) \left[ 2 g^{+-zz} ( \omega^{\prime} , - \omega^{\prime} , \omega , - \omega ) +   g^{zzzz} ( \omega^{\prime} , - \omega^{\prime} , \omega , - \omega ) \right]
 \nonumber
 \\
 &-  \frac{2}{\beta N} \sum_{\bd{q}} J^2_{ \bd{q} } J_{ \bd{q} + \bd{k} } 
 \sum_{\omega^{\prime}} g ( \omega^{\prime} ) g^{ +-z} ( \omega^{\prime} , - \omega 
 - \omega^{\prime} , \omega ) g^{+-z} ( \omega + \omega^{\prime}, - \omega^{\prime} , - \omega ).
 \label{eq:Pi3int}
 \end{align}
It is convenient to perform the 
frequency sums for finite field and then take the limit $ \beta H \rightarrow 0$ using the
identities
 \begin{align}
 &  \lim_{ H \rightarrow 0} 
 \frac{1}{\beta^2} 
 \sum_{\omega_1 \omega_2} 
 \Bigl[
 g^{+-zz} ( \omega_1 , \omega_2 , \omega , - \omega_1 - \omega_2 - \omega)
 g^{+-zz} ( - \omega_2 , - \omega_1 , - \omega, \omega_1 + \omega_2 + \omega )  \nonumber
 \\
 &  \hspace{18mm} + \frac{1}{6}  g^{zzzz} ( \omega_1 , \omega_2 , \omega , - \omega_1 - \omega_2 - \omega)
 g^{zzzz} ( - \omega_2 , - \omega_1 , - \omega, \omega_1 + \omega_2 + \omega ) 
 \Bigr] 
 \nonumber
 \\
 = {} & 
 \delta_{\omega,0} \frac{\beta^4}{72} \left[ b_1^2 + 20 b_3^2 \right]
 + ( 1 - \delta_{\omega , 0 } ) \frac{ \beta^2 b_1^2 }{2 \omega^2 },
 \end{align}
and
 \begin{align}
 & 
 \lim_{ H \rightarrow 0} \frac{1}{\beta^2} \sum_{\omega_1 \omega_2}
 g^{+-zzz} ( \omega_1 , \omega_2 , - \omega_1 - \omega_2 , \omega , - \omega )
 g^{+-z} ( - \omega_2 , - \omega_1 , \omega_1 + \omega_2 )
 \nonumber
 \\
 = {} &  \delta_{\omega ,0} \frac{\beta^4}{72} \left[ b_1^2 - 10 b_1 b_3 \right] 
 - ( 1 - \delta_{\omega,0} ) \frac{\beta^2 b_1^2 }{ 2 \omega^2 }.
 \end{align}

Finally, let us give some technical details for the derivation of our second-order result for
chiral non-linear susceptibility $G^{+-z (2)} ( K_1 , K_2 , K_3 )$
in Eq.~\eqref{eq:G2chiral}.
All diagrams contributing up to second order in the exchange couplings
are shown in Fig.~\ref{fig:G3pert}.
Assuming $H=0$ and that none of the frequencies
$\omega_1$, $\omega_2$, and $\omega_3$ vanishes, only four second-order diagrams shown in
 Fig.~\ref{fig:G3pert} contribute: the diagram where the five-spin block is connected with the
two-spin block in the third line, and the three diagrams
where the four-spin block is connected  with the three-spin block 
in the fourth line of Fig.~\ref{fig:G3pert}. The contribution from the diagram containing the five-spin block is
 \begin{equation}
 G^{+-z(2a)} ( K_1 , K_2 , K_2 ) =
 \frac{1}{2 \beta N} \sum_{\bd{q}} J^2_{\bd{q}} \sum_{\omega}
 g ( \omega ) \left[ 2 g^{++--z} ( \omega_1, \omega , \omega_2 ,  - \omega , \omega_3 )
 + g^{+-zzz} ( \omega_1 , \omega_2 , \omega_3 , \omega , - \omega ) \right].
 \end{equation}
For $H \rightarrow 0$ we may replace $g ( \omega ) \rightarrow  \beta b_1 \delta_{\omega , 0 }$ so that the frequency sum can be carried out trivially. 
Using our results for the five-spin blocks given in
Eqs.~\eqref{eq:G5a} and \eqref{eq:G5b}
we obtain for $H \rightarrow 0$,
 \begin{equation}
  G^{+-z (2a)} ( K_1 , K_2 , K_2 ) =  i \beta b_1^2 \left( \frac{1}{N} \sum_{\bd{q}} J_{\bd{q}}^2
 \right) \left[  \frac{2}{\omega_1 \omega_2} 
 \left( \frac{1}{\omega_1} - \frac{1}{\omega_2} \right) - \frac{1}{\omega_1 \omega_2 \omega_3} - \frac{2}{ \omega_2 \omega_3^2 } \right].
 \end{equation}
Finally, using the constraint $\omega_1 + \omega_2 + \omega_3 =0$
we may write the frequency-dependent term in a more symmetric form,
 \begin{align}
   \frac{2}{\omega_1 \omega_2} 
 \left( \frac{1}{\omega_1} - \frac{1}{\omega_2} \right) - \frac{1}{\omega_1 \omega_2 \omega_3} - \frac{2}{ \omega_2 \omega_3^2 } & = \frac{ ( \omega_1 - \omega_2 )
 ( \omega_2 - \omega_3 ) ( \omega_3 - \omega_1 )}{ \omega_1^2 \omega_2^2 \omega_3^2}
 \nonumber
 \\
 & = \frac{1}{\omega_1^2} \left( \frac{1}{\omega_2} - \frac{1}{\omega_3} \right)
  + \frac{1}{\omega_2^2} \left( \frac{1}{\omega_3} - \frac{1}{\omega_1} \right)
 + \frac{1}{\omega_3^2} \left( \frac{1}{\omega_1} - \frac{1}{\omega_2} \right),
 \end{align}
which gives  
the momentum-independent contribution to 
$G^{xyz (2)} ( K_1 , K_2, K_3 )$ in
Eq.~\eqref{eq:G2chiral}. The momentum-dependent part involving the form factors 
$\lambda_{\bd{k}}$ defined in Eq.~\eqref{eq:fkdef} arises from the three diagrams 
in the fourth row of Fig.~\ref{fig:G3pert}
where
the four-spin block is connected to the three-spin block. 
 The explicit evaluation of these diagrams is 
straightforward. To write the resulting frequency dependence in the form
given in Eqs.~\eqref{eq:G2chiral} and \eqref{eq:Gxyz_complete} below, we use again some identities implied
by frequency conservation such as
 \begin{equation}
 \frac{2}{ \omega_1 \omega_3^2} + \frac{1}{\omega_1 \omega_2 \omega_3} =
 \frac{1}{\omega_3^2} \left( \frac{1}{\omega_1} - \frac{1}{\omega_2} \right).
 \end{equation}
The evaluation of the diagrams in Fig.~\ref{fig:G3pert} for the chiral three-point function if one of the three frequencies vanishes proceeds analogously.
The resultant high-temperature expansion of the chiral non-linear susceptibility $ G^{ x y z} = -i G^{ + - z } $ of a spin-$S$ Heisenberg magnet is
in the limit $ H \to 0 $ given by
\begin{align}
G^{ x y z }_{ i_1 i_2 i_3 } ( \omega_1 , \omega_2 , \omega_3 )
= {} &
\beta b_1
\frac{ \delta_{ \omega_1 , 0 } }{ \omega_2 }
\left( 1 - \delta_{ \omega_2 , 0 } \right)
\left\{
\delta_{ i_1 i_2 } \delta_{ i_2 i_3 } \left[
1 + 5 \sum_j J_{ i_1 j }^2 \left(
\frac{ \beta^2 b_3 }{ 6 } - \frac{ b_1 }{ \omega_2^2 } 
\right)
\right]
+ 3 \left( 
\delta_{ i_3 i_1 } J_{ i_1 i_2 }^2 +
\delta_{ i_1 i_2 } J_{ i_2 i_3 }^2 
\right) \frac{ b_1 }{ \omega_2^2 }
\right.
\nonumber\\
& \left. \phantom{  
\beta b_1
\frac{ \delta_{ \omega_1 , 0 } }{ \omega_2 }
\left( 1 - \delta_{ \omega_2 , 0 } \right)
\Biggl\{
}
- \delta_{ i_2 i_3 } b_1 \left[
\beta  J_{ i_3 i_1 }  + J_{ i_3 i_1 }^2 b_1 \left(
\frac{ \beta^2 }{ 12 } + \frac{ 1 }{ \omega_2^2 }
\right)
- \beta^2 \sum_j J_{ i_3 j } J_{ j i_1 } b_1  
\right]
\right\}
\nonumber\\
&
+ 
\left( 1 - \delta_{ \omega_1 , 0 } \right)
\left( 1 - \delta_{ \omega_2 , 0 } \right)
\left( 1 - \delta_{ \omega_3 , 0 } \right)
\left(
\delta_{ i_1 i_2 } \delta_{ i_2 i_3 } \sum_j J_{ i_1 j }^2 
- \delta_{ i_2 i_3 } J_{ i_3 i_1 }^2
\right)
\frac{ \beta b_1^2 }{ \omega_1^2 } \left(
\frac{ 1 }{ \omega_2 } - \frac{ 1 }{ \omega_3 }
\right)
\nonumber\\
& 
+ \text{cyclic permutations of 
$\left\{ (i_1,\omega_1), (i_2,\omega_2), (i_3,\omega_3) \right\}$}
+ {\cal O} ( J^3 ).
\label{eq:Gxyz_complete}
\end{align}
\end{widetext}

\section{The Heisenberg trimer}

\label{app:trimer}

\renewcommand{\theequation}{C\arabic{equation}}

\setcounter{equation}{0}
\setcounter{subsection}{0}

To test the reliability and overall usefulness of series expansion methods,
it is desirable to benchmark them against exactly solvable (toy) models.
To that end,
we here consider the Heisenberg trimer with Hamiltonian
\begin{equation} \label{eq:trimer}
{\cal H} = J \left(
\bm{S}_1 \cdot \bm{S}_2 +
\bm{S}_2 \cdot \bm{S}_3 +
\bm{S}_3 \cdot \bm{S}_1 
\right) ,
\end{equation}
where $ \bm{S}_i $, $ i \in \{ 1 , 2 , 3 \} $, are spin-$ 1/2 $ operators and $ J > 0 $ is an antiferromagnetic exchange coupling.
Note that the trimer \eqref{eq:trimer} is essentially a one-dimensional chain of $ N = 3 $ spins with periodic boundary conditions.
Since it can be straightforwardly diagonalized in the basis of the total spin $ \bm{S}_1 + \bm{S}_2 + \bm{S}_3 $,
we can use it to test the high-temperature series expansions for spin correlation functions presented in this paper
against exact solutions.
For example,
the free energy,
as well as the local (11) and non-local (12) dynamic spin susceptibility
and their respective third-order high-temperature series are given by
\begin{subequations} \label{eq:F_trimer}
\begin{align}
F 
& = - \frac{ 1 }{ \beta } \ln \left[
8 \cosh \left( 3 \beta J / 4 \right)
\right]
\\
& = - \frac{ 3 }{ \beta } \ln 2 
- \frac{ 9 \beta J^2 }{ 32 }
+ {\cal O} ( J^4 ) ,
\end{align}
\end{subequations}
\begin{subequations} \label{eq:G2_trimer_local}
\begin{align}
G_{ 1 1 } ( \omega ) 
& = \beta \delta_{ \omega , 0 } \frac{ 5 }{ 36 }
+ \frac{ 4 J \tanh \left( 3 \beta J / 4 \right) }{
3 \left( 9 J^2 + 4 \omega^2 \right) }
\\
& = 
\beta \delta_{ \omega , 0 } \left( 
\frac{ 1 }{ 4 } - \frac{ \beta^2 J^2 }{ 48 } \right)
+ \left( 1  - \delta_{ \omega , 0 } \right) 
\frac{ \beta J^2 }{ 4 \omega^2 }
+ {\cal O}( J^4 )  ,
\end{align}
\end{subequations}
and
\begin{subequations} \label{eq:G2_trimer_non-local}
\begin{align}
G_{ 1 2 } ( \omega )
= {} &
\frac{ \beta \delta_{ \omega , 0 } }{ 36 } \left[
-2 + 3 \tanh \left( 3 \beta J / 4 \right) 
\right]
\nonumber\\
& - \left( 1 - \delta_{ \omega , 0 } \right)
\frac{ 2 J \tanh \left( 3 \beta J / 4 \right) }{
3 \left( 9 J^2 + 4 \omega^2 \right) }
\\
= {} &
\beta \delta_{ \omega , 0 } \left(
- \frac{ \beta J }{ 16 } + \frac{ \beta^2 J^2 }{ 96 } + \frac{ 3 \beta^3 J^3 }{ 256 }
\right)
\nonumber\\
& - \left( 1  - \delta_{ \omega , 0 } \right) 
\frac{ \beta J^2 }{ 8 \omega^2 }
+ {\cal O} ( J^4 ) .
\end{align}
\end{subequations}
Given these perturbation series for the susceptibility,
one can also test the accuracy of resummations based on the interaction-irreducible part $ \Pi $
introduced in Eq.~\eqref{eq:GPi}.
Expanding $ \Pi $ itself to order $ J^3 $ yields the resummation
\begin{widetext}
\begin{subequations} \label{eq:G2_Pi_trimer}
\begin{align}
G^{ ( \Pi ) }_{ 1 1 } ( \omega )
& =
\frac{ \beta \delta_{ \omega , 0 } }{ 6 }
\left[
\frac{
4 ( \beta J - 4 )^2 ( \beta J + 2 )
}{
128 - \beta J ( \beta J - 4 )^2 ( \beta J + 2 )
}
+ \frac{
32 + ( \beta J )^2 ( \beta J + 8 )
}{
64 + 32 \beta J - 8 ( \beta J )^3 - ( \beta J )^4
}
\right]
+ \left( 1 - \delta_{ \omega , 0 } \right)
\frac{ 2 \beta J^2 }{
8 \omega^2 - 3 \beta J^3
} ,
\\
G^{ ( \Pi ) }_{ 1 2 } ( \omega )
& =
\frac{ \beta \delta_{ \omega , 0 } }{ 6 }
\left[
- \frac{
2 ( \beta J - 4 )^2 ( \beta J + 2 )
}{
128 - \beta J ( \beta J - 4 )^2 ( \beta J + 2 )
}
+ \frac{
32 + ( \beta J )^2 ( \beta J + 8 )
}{
64 + 32 \beta J - 8 ( \beta J )^3 - ( \beta J )^4
}
\right]
- \left( 1 - \delta_{ \omega , 0 } \right)
\frac{ \beta J^2 }{
8 \omega^2 - 3 \beta J^3
} .
\end{align}
\end{subequations}
If one on the other hand expands the \emph{inverse} interaction-irreducible susceptibility $ \Pi^{ - 1 } $ to order $ J^3 $,
one obtains instead the resummation
\begin{subequations} \label{eq:G2_Sigma_trimer}
\begin{align}
G^{ ( \Pi^{ - 1 } ) }_{ 1 1 } ( \omega )
& =
\frac{ 8 \beta \delta_{ \omega , 0 } }{ 3 }
\left[
\frac{
2
}{
32 - \beta J ( \beta J - 4 ) ( \beta J - 2 )
}
+ \frac{
1
}{
32 + \beta J ( \beta J + 4 )^2
}
\right]
+ \left( 1 - \delta_{ \omega , 0 } \right)
\frac{ 2 \beta J^2 }{
8 \omega^2 - 3 \beta J^3
} ,
\\
G^{ ( \Pi^{ - 1 } ) }_{ 1 2 } ( \omega )
& =
\frac{ 8 \beta \delta_{ \omega , 0 } }{ 3 }
\left[
- \frac{
1
}{
32 - \beta J ( \beta J - 4 ) ( \beta J - 2 )
}
+ \frac{
1
}{
32 + \beta J ( \beta J + 4 )^2
}
\right]
- \left( 1 - \delta_{ \omega , 0 } \right)
\frac{ \beta J^2 }{
8 \omega^2 - 3 \beta J^3
} ,
\end{align}
\end{subequations}
Note that while both resummation schemes give the same dynamic behavior to this order,
the static limit $ \omega \to 0 $ is extrapolated very differently.
The exact formulas for the free energy and spin susceptibilities are compared
with their third-order expansions and resummations in Fig.~\ref{fig:g2_trimer}.
\begin{figure*}[tb]
 \begin{center}
  \centering
\vspace{7mm}
 \includegraphics[width=.95\textwidth]{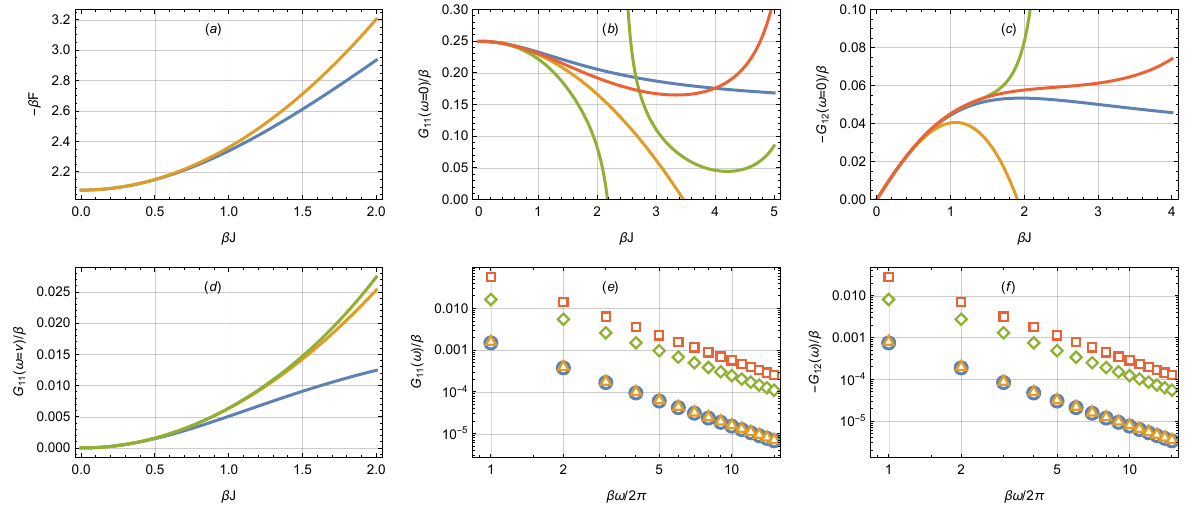}
   \end{center}
  \caption{%
Comparison of exact trimer correlation functions and their order $ J^3 $ high-temperature series.
(a) shows the free energy $ F $,
while (b)-(f) display the dynamic spin susceptibility $ G_{ i j } ( \omega ) $ and its resummations.
In (a)-(d),
blue and orange lines are exact and order $ J^3 $ perturbative results given in Eqs.~\eqref{eq:F_trimer}-\eqref{eq:G2_trimer_non-local},
whereas green and red lines are the resummations \eqref{eq:G2_Pi_trimer} and \eqref{eq:G2_Sigma_trimer} based on $ \Pi $ and $ \Pi^{-1} $, respectively.
(b) and (c) show the static limit $ \omega \to 0 $ of $ G_{ i j } ( \omega ) $,
and (d) the value of the local susceptibility $ G_{ 1 1 } ( \omega ) $ at the first finite Matsubara frequency $ \omega = \nu = 2 \pi / \beta $.
(e) and (f) are the Matsubara frequency dependence of $ G_{ i j } ( \omega ) $ on a double-logarithmic scale,
where blue circles and green diamonds are exact values for $ \beta J = 1 /2 $ and $ \beta J = 3 $, respectively,
and orange triangles and red squares are the corresponding third-order high-temperature expansions.
We do not show the $ \Pi $ and $ \Pi^{-1} $ resummations in (e) and (f) because they lie on top of the perturbation series for these values of $ \beta J $. 
}
\label{fig:g2_trimer}
\end{figure*}
One can see that the order $ J^3 $ perturbation series is surprisingly accurate up to $ \beta J \approx 1 $.
Moreover,
both the $ \Pi $ and the $ \Pi^{-1} $ resummations schemes,
\eqref{eq:G2_Pi_trimer} and \eqref{eq:G2_Sigma_trimer}, respectively,
introduce artificial divergences when extrapolated beyond the high-temperature regime $ \beta J < 1 $.
From the analytical formula \eqref{eq:G2_Pi_trimer} for the $ \Pi $ resummation,
one finds that for $ \omega = 0 $, 
there are two unphysical poles at $ \beta J \approx 2.38 $ and $ 5.71 $,
which are clearly visible in Figs.~\ref{fig:g2_trimer} (b) and (c).
Since the first pole is at relatively high temperatures,
this leads to the curious result that the resummation based on $ \Pi $ actually performs \emph{worse} 
than the bare perturbation series for $ G_{ 1 1 } ( \omega = 0 ) $.
The $ \Pi^{-1} $ resummation \eqref{eq:G2_Sigma_trimer} on the other hand
yields a comparatively accurate and qualitatively correct extrapolation up to $ \beta J \approx 3 $.
This is overall consistent with our findings from Sec.~\ref{sec:dynamic_spin_susceptibility},
where the extrapolation based on $ \Pi^{-1} $ allowed us to estimate critical temperatures of Heisenberg magnets,
which lie roughly in the regime $  \beta J = {\cal O} ( 1 ) $ for $ S = 1 / 2 $,
while the extrapolation based on $ \Pi $ failed.
However,
for even lower temperatures,
the $ \Pi^{-1} $ resummation \eqref{eq:G2_Sigma_trimer} also develops an unphysical singularity at $ \beta J \approx 5.59 $. 
At a slightly higher temperature
$ \beta J = 2 ( 4 \pi^2 / 3 )^{ 1 / 3 } \approx 4.72 $
the dynamic ($ \omega \neq 0 $) part of both resummations likewise has an unphysical pole at the first finite Matsubara frequency.
Despite this,
Figs.~\ref{fig:g2_trimer} (e) and (f) show that the power law frequency dependence of the spin susceptibility 
is actually captured very well already by the leading order in perturbation theory,
even at low temperatures $ \beta J > 1 $ where it is uncontrolled. 
This goes some way towards explaining why it is sufficient to retain only the leading frequency dependence of dynamic spin correlations
and still obtain good results for the phase diagram of various quantum Heisenberg models with the spin FRG,
provided that one uses a sufficiently sophisticated $ \Pi^{-1} $ resummation for the static spin correlations \cite{Tarasevych22b,Rueckriegel24}.

\begin{figure*}[tb]
 \begin{center}
  \centering
\vspace{7mm}
 \includegraphics[width=.95\textwidth]{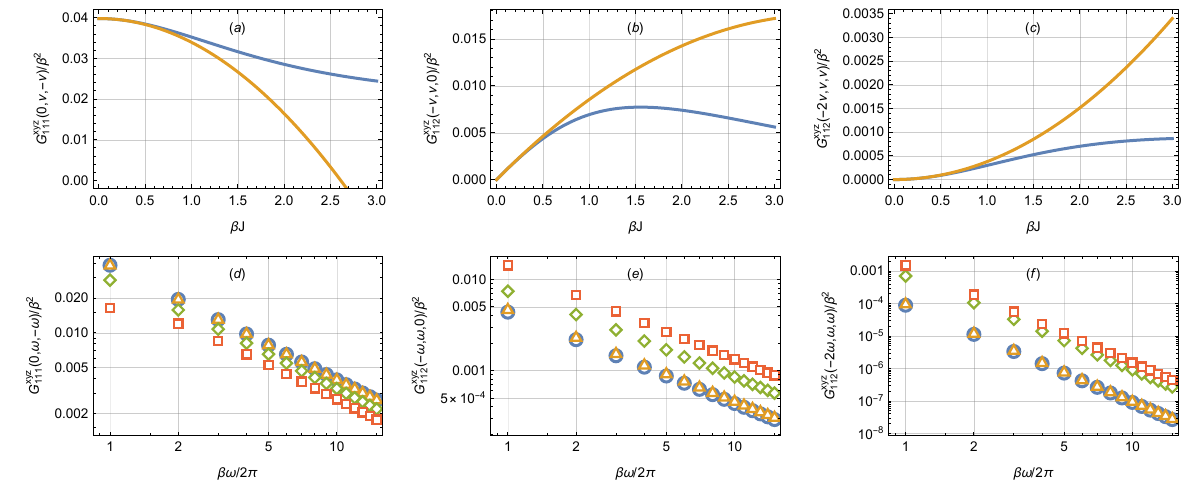}
   \end{center}
  \caption{%
Comparison of the exact chiral non-linear susceptibility $ G^{ x y z }_{ i j k } ( \omega_1 , \omega_2 , \omega_3 ) $ of the Heisenberg trimer
with the order $ J^2 $ perturbation series;
see Eqs.~\eqref{eq:G3_trimer_111} and \eqref{eq:G3_trimer_112}.
In (a)-(c),
blue lines are exact and orange lines are the respective second-order perturbation series,
where  $ \nu = 2 \pi / \beta $ denotes the first finite Matsubara frequency.
(d)-(f) are the Matsubara frequency dependence of the chiral non-linear susceptibility on a double-logarithmic scale,
where blue circles and green diamonds are exact values for $ \beta J = 1 /2 $ and $ \beta J = 2 $, respectively,
and orange triangles and red squares are the corresponding second-order high-temperature expansions.
Note that (c) and (f) correspond to the second-harmonic generation discussed in Sec.~\ref{sec:chiral_non-linear_susceptibility},
which is only possible in the trimer if exactly two spins are on the same site [see Eqs.~\eqref{eq:G3_trimer_111}-\eqref{eq:G3_trimer_123}].
}
\label{fig:g3_trimer}
\end{figure*}
Finally,
let us consider the exact chiral non-linear susceptibility $ G^{ x y z }_{ i j k } ( \omega_1 , \omega_2 , \omega_3 ) $ of the trimer
and its second-order high-temperature expansion \eqref{eq:Gxyz_complete}.
If all three spins are on the same site,
it is given by
\begin{subequations} \label{eq:G3_trimer_111}
\begin{align}
G^{ x y z }_{ 1 1 1 } ( \omega_1 , \omega_2 , \omega_3 )
= {} &
\frac{ 32
( \omega_1 - \omega_2 ) 
( \omega_2 - \omega_3 )
( \omega_3 - \omega_1 )
J \tanh \left( 3 \beta J / 4 \right)
}{
3
\left( 9 J^2 + 4 \omega_1^2 \right)
\left( 9 J^2 + 4 \omega_2^2 \right)
\left( 9 J^2 + 4 \omega_3^2 \right)
}
\nonumber\\
&
+ \Biggl\{ 
\beta \delta_{ \omega_1 , 0 }
\left( 1 - \delta_{ \omega_2 , 0 } \right) 
\left[
\frac{ 1 }{ 12 \omega_2 }
+ \frac{ 
2 \omega_2^2 
+ 3 J^2 \tanh \left( 3 \beta J / 4 \right)
}{ 9 \omega_2 \left( 9 J^2 + 4 \omega_2^2 \right) }
+ \frac{ 
64 \omega_2^3 \tanh \left( 3 \beta J / 4 \right)
}{ 27 \beta J \left( 9 J^2 + 4 \omega_2^2 \right)^2 }
\right]
\nonumber\\
& \phantom{aaaa} + \text{cyclic permutations of 
$ \{ \omega_1 , \omega_2 , \omega_3 \} $}
\Biggr\}
\\
= {} &
\left( 1 - \delta_{ \omega_1 , 0 } \right)
\left( 1 - \delta_{ \omega_2 , 0 } \right)
\left( 1 - \delta_{ \omega_3 , 0 } \right)
\frac{ 
( \omega_1 - \omega_2 ) 
( \omega_2 - \omega_3 )
( \omega_3 - \omega_1 )
}{ 8 \omega_1^2 \omega_2^2 \omega_3^2 } \beta J^2
\nonumber\\
& + \left\{ \beta \delta_{ \omega_1 , 0 }
\left( 1 - \delta_{ \omega_2 , 0 } \right) \left[
\frac{ 1 }{ 4 \omega_2 } 
\left( 1 - \frac{ \beta^2 J^2 }{ 12 } \right) 
- \frac{ 5 J^2 }{ 8 \omega_2^3 }
\right]
+ \text{cyclic permutations of 
$ \{ \omega_1 , \omega_2 , \omega_3 \} $}
\right\} 
%
%
+ {\cal O} ( J^3 ).
\end{align}
\end{subequations}
If only two spins are on the same site,
one finds instead
\begin{subequations} \label{eq:G3_trimer_112}
\begin{align}
G^{ x y z }_{ 1 1 2 } ( \omega_1 , \omega_2 , \omega_3 )
= {} &
\frac{
8 ( \omega_1 - \omega_2 ) 
( 2 \omega_1 \omega_2 - 9 J^2 )
J \tanh \left( 3 \beta J / 4 \right)
}{
3 
\left( 9 J^2 + 4 \omega_1^2 \right)
\left( 9 J^2 + 4 \omega_2^2 \right)
\left( 9 J^2 + 4 \omega_3^2 \right)
}
\nonumber\\
& 
+ \beta \delta_{ \omega_1 , 0 }
\left( 1 - \delta_{ \omega_2 , 0 } \right) J^2
\frac{
3 - 2 \tanh \left( 3 \beta J / 4 \right)
}{
12 \omega_2 \left( 9 J^2 + 4 \omega_2^2 \right)
}
+ \beta \delta_{ \omega_2 , 0 }
\left( 1 - \delta_{ \omega_3 , 0 } \right) J^2
\frac{
3 - 2 \tanh \left( 3 \beta J / 4 \right)
}{
12 \omega_3 \left( 9 J^2 + 4 \omega_3^2 \right)
}
\nonumber\\
& + \beta \delta_{ \omega_3 , 0 }
\left( 1 - \delta_{ \omega_1 , 0 } \right)
\left[
\frac{ 1 }{ 36 \omega_1 } 
+ \frac{ \omega_1 }{ 9 \left( 9 J^2 + 4 \omega_1^2 \right) }
- \frac{
( J^2 + 2 \omega_1^2 ) \tanh \left( 3 \beta J / 4 \right)
}{
6 \omega_1 \left( 9 J^2 + 4 \omega_1^2 \right)
}
\right]
\\
= {} &
\left( 1 - \delta_{ \omega_1 , 0 } \right)
\left( 1 - \delta_{ \omega_2 , 0 } \right)
\left( 1 - \delta_{ \omega_3 , 0 } \right)
\frac{ \omega_1 - \omega_2 
}{ 16 \omega_1 \omega_2 \omega_3^2 } \beta J^2
+ \beta \delta_{ \omega_1 , 0 }
\left( 1 - \delta_{ \omega_2 , 0 } \right)
\frac{ 3 J^2 }{ 16 \omega_2^3 }
+ \beta \delta_{ \omega_2 , 0 }
\left( 1 - \delta_{ \omega_3 , 0 } \right)
\frac{ 3 J^2 }{ 16 \omega_3^3 }
\nonumber\\
& + \beta \delta_{ \omega_3 , 0 }
\left( 1 - \delta_{ \omega_1 , 0 } \right)
\left[
\frac{ \beta J }{ 16 \omega_1 } 
\left( - 1 + \frac{ \beta J }{ 6 } \right)
- \frac{ J^2 }{ 16 \omega_1^3 }
\right]
+ {\cal O} ( J^3 ) .
\end{align}
\end{subequations}
If all spins are on different sites,
the chiral non-linear susceptibility must be of order $ J^3 $,
\begin{equation} \label{eq:G3_trimer_123}
G^{ x y z }_{ 1 2 3 } ( \omega_1 , \omega_2 , \omega_3 )
= - \beta \delta_{ \omega_1 , 0}
\left( 1 - \delta_{ \omega_2 , 0 } \right) 
\frac{ 
5 J^2 \tanh \left( 3 \beta J / 4 \right) 
}{
12 \omega_2 \left( 9 J^2 + 4 \omega_2^2 \right)
}
+ \text{cyclic permutations of 
$ \{ \omega_1 , \omega_2 , \omega_3 \} $}
= {\cal O} ( J^3 ) .
\end{equation}
Note that this last expression is rather less involved than either \eqref{eq:G3_trimer_111} or \eqref{eq:G3_trimer_112},
owing to the fact that spins on different sites commute.
Exact formulas and perturbation expansions of the chiral non-linear susceptibility are compared in Fig.~\ref{fig:g3_trimer}.
The results are similar to the ones for the dynamic spin susceptibility in Figs.~\ref{fig:g2_trimer} (d)-(f),
with the second-order perturbation theory failing a bit earlier, around $ \beta J \approx 0.6 $-$ 0.8 $.
Again,
the frequency dependence is described very well by a power law dependence for arbitrary $ \beta J $ 
that is consistent with the leading order in perturbation theory in $ \beta J $.
We trace this back to the fact that the non-trivial frequency dependence of the spin correlations 
is generated by the commutation relations of spins at the same site and hence dominated by local spin correlations,
corresponding to the leading order in the exchange coupling $ J $.
\end{widetext}


\begin{thebibliography}{99}
\bibitem{footnote_nearest_neighbor}
Note that Eq.~\eqref{eq:gamma3_sum} only holds in the thermodynamic limit $ N \to \infty $,
while the other two relations of Eqs.~\eqref{eq:gamma_sums} also hold for finite systems.
\end{thebibliography}
\end{document}